\documentclass[%
 reprint,
 superscriptaddress,
 amsmath,amssymb,
 aps,
prx,
twocolumn
]{revtex4-1}

\usepackage{bibunits}
\usepackage{float}
\usepackage{graphicx, afterpage}
\usepackage{kantlipsum}     
\usepackage{graphicx}
\usepackage{dcolumn}
\usepackage{bm}

\defaultbibliography{reff}
\defaultbibliographystyle{apsrev4-1}

\begin{document}

\title{Transport evidence for three-dimensional topological superconductivity in doped $\beta$-PdBi$_2$}

\author{Ayo Kolapo}
\affiliation{Texas Center for Superconductivity and Department of Physics, University of Houston, 3201 Cullen Boulevard, Houston, Texas 77204, USA}
\author{Tingxin Li} 
\affiliation{Department of Physics and Astronomy, Rice University, Houston, Texas 77251, USA}
\author{Pavan Hosur}
\affiliation{Texas Center for Superconductivity and Department of Physics, University of Houston, 3201 Cullen Boulevard, Houston, Texas 77204, USA}
\author{John H. Miller} 
\affiliation{Texas Center for Superconductivity and Department of Physics, University of Houston, 3201 Cullen Boulevard, Houston, Texas 77204, USA}

\date{\today}

\begin{abstract}
Interest in topological states of matter exploded over a decade ago with the theoretical prediction and experimental detection of three-dimensional topological insulators, especially in bulk materials that can be tuned out of it by doping. However, their superconducting counterpart, the time-reversal invariant three-dimensional topological superconductor, has evaded discovery thus far. In this work, we provide transport evidence that K-doped $\beta$-PdBi$_2$ is a 3D
time-reversal-invariant topological superconductor. In particular, we find signatures of Majorana surface states protected by time-reversal symmetry\textemdash hallmark of this phase\textemdash in soft point-contact spectroscopy, while the bulk system shows signatures of odd-parity pairing via upper-critical field and magnetization measurements. Odd-parity pairing can be argued, using existing knowledge of the band structure of $\beta$-PdBi$_2$, to result in 3D topological superconductivity. Moreover, we find that the undoped system is a trivial superconductor. Thus, we discover $\beta$-PdBi$_2$ as a unique material that, on doping, can potentially undergo an unprecedented topological quantum phase transition in the superconducting state.

\begin{description}
\item[Subject Areas:]
Topological Superconductors, Topological Insulators
\end{description}

\end{abstract}

\maketitle


\begin{bibunit}

\section{\label{sec:level1}INTRODUCTION}

According to the traditional Landau-Ginzburg paradigm, states of matter are defined by the symmetries broken in thermal equilibrium that are preserved by the underlying Hamiltonian, and phase transitions acquire universal features that only depend on the symmetries involved and the spatial dimension. However, this definition proves inadequate for topological phases, in which the ground state wavefunction of the bulk system is characterized by a global, topological quantum number which distinguishes it from a conventional phase with the same symmetries \cite{moore2010birth, schnyder2008classification, hasan2010colloquium}. Naturally, critical points separating these phases fall outside the tradition paradigm as well. The most striking consequence of the non-trivial bulk topology is the presence of robust surface states where the bulk terminates. One of the most celebrated topological phases in condensed matter systems is the time-reversal symmetric strong topological insulator (TI) in three dimensions, which is characterized by a $\mathbb{Z}_2$ topological invariant $\nu=\text{odd/even}$ \cite{fu2007topological, chen2009experimental}. The surface manifestation of the bulk topology in this phase is the presence of an odd number of pseudo-relativistic, helical surface states that are robust against non-magnetic perturbations. Numerous materials have been predicted to be in this phase, and many of them have been experimentally confirmed. Additionally, several TIs can be tuned into trivial insulators with doping, thus allowing experimental access to the quantum critical point separating them.

A close cousin of the topological insulator is the time-reversal symmetric topological superconductor (TSC) in 3D \cite{schnyder2008classification}. (See Fig. 1). Here, the superconducting gap plays the role of the insulating gap of the insulator, the topological invariant is $\nu\in\mathbb{Z}=0,1,2\dots$, and the surface hosts $\nu$ helical Majorana fermions instead of electrons. These states have been shown to serve as a condensed matter realization of spontaneous supersymmetry breaking \cite{qi2009time,grover2014emergent}, while a related particle -- the Majorana zero mode, shown to exist in various heterostructures of superconductors and spin-orbit coupled systems including topological insulators \cite{mourik2012signatures, fu2008superconducting} -- has been extensively proposed as the building block for a fault-tolerant quantum computer \cite{kitaev2003fault}. Despite such remarkable theoretical predictions, however, 3D topological superconductors have been elusive experimentally. The main material platform that has been studied experimentally is the prototypical TI Bi$_2$Se$_3$ doped with Cu, in addition to theoretical predictions on a handful of rare earth compounds \cite{chadov2010tunable}. Unfortunately, undoped Bi$_2$Se$_3$ does not display superconductivity at ambient pressures, so a topological-to-trivial superconductor phase transition does not occur in this system.

\begin{figure*}[t!]
\centering
\includegraphics[width=0.7\textwidth,,height=7cm]{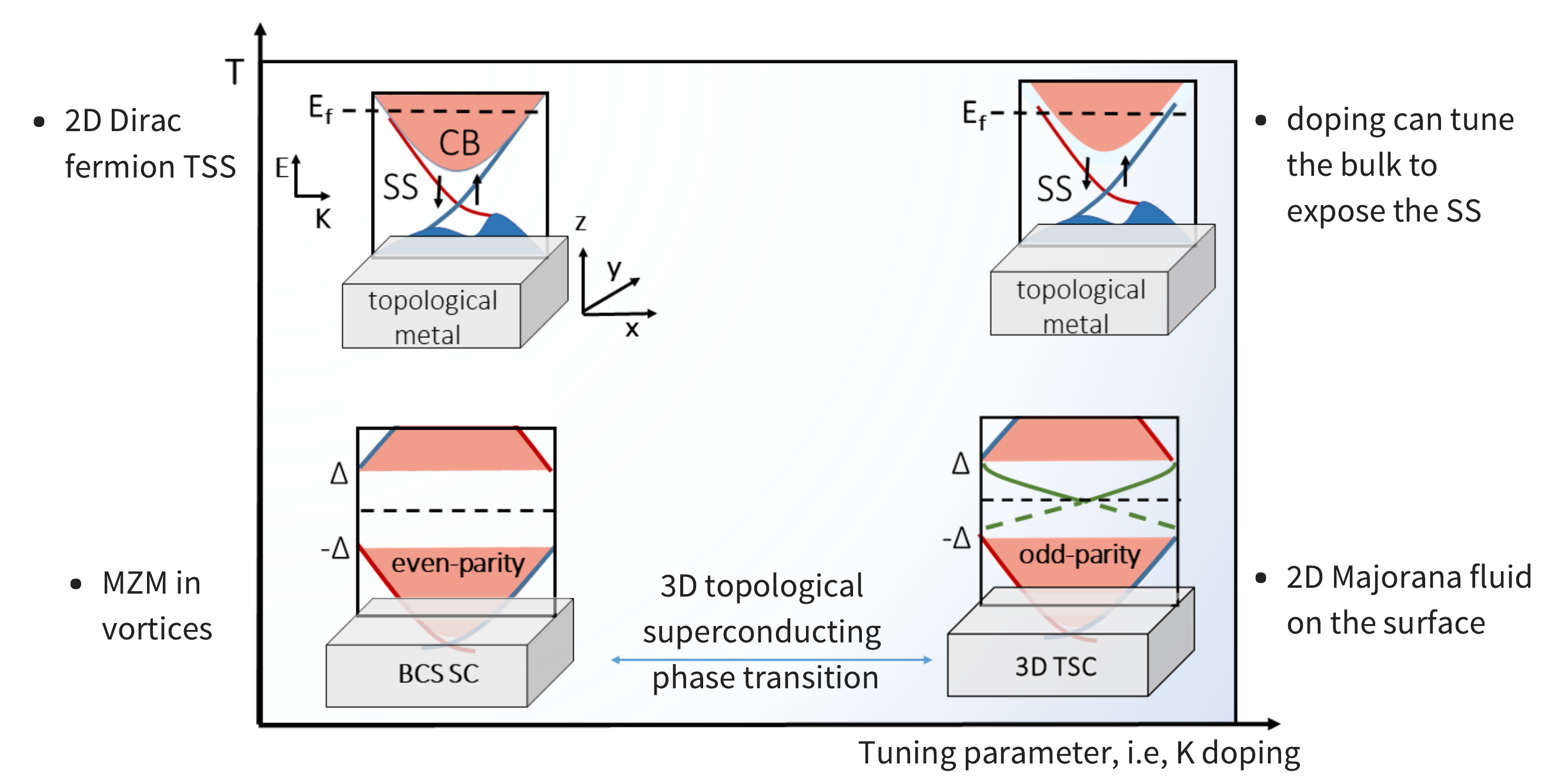}
\caption{ \textbf{3D topological superconducting phase transition. } Upper-left: topological metal: Fermi level E$_f$ is placed above (or below) the topological surface state SS. Upper-right: Doping drains the bulk bands and lowers (raises) the Fermi level to expose the SS. Lower-left: Inducing even-parity superconductivity in the bulk of a topological metal gaps out the bulk and the Dirac SS, while (lower-right) odd-parity paring turns the bulk system into a 3D topological superconductor, which hosts in-gap, helical Majorana fermion surface states. While signatures of Majarana zero mode (MZM) have been demonstrated in several experiments (in superconducting TI and proximity induced superconductivity heterostructure), the 2D Majorana fluid unique to 3D TSC has been elusive. Importantly, this system could exhibit a topological quantum phase transition in the superconducting state. } 
\end{figure*}

In this work, we investigate the transport properties of $\beta$-PdBi$_2$ single crystals with and without K doping. Our normal state magneto-transport experiments on K-doped $\beta$-PdBi$_2$ show that the longitudinal magnetoresistivity (LMR) exhibits weak anti-localization (WAL) at low temperatures, which is a signature of the presence of spin-momentum locked Dirac fermions, while the undoped system lacks any such features in LMR. Next, we probe superconductivity in these systems using various bulk and surface transport measurements. We find the undoped material to be a topologically trivial superconductor; in contrast, we find evidence that the doped system realizes a TSC. Specifically, point-contact spectroscopy (PCS) in the superconducting state shows signatures of Majorana surface states protected by time-reversal symmetry as theoretically predicted for time-reversal-invariant 3D topological superconductors. On studying the superconductivity in the bulk, we find that (i) the upper critical field exceeds the prediction by the Werthemer-Helfand-Hohenberg (WHH) orbital model for conventional $s$-wave pairing, but is consistent with the prediction for polar $p$-wave pairing, and (ii) the magnetization shows an anomalous behavior, characteristic of spin-triplet superconductivity. Fu and Berg showed that time-reversal symmetric odd-parity superconductors are topological if their normal state Fermi surface encloses an odd number of time-reversal invariant momenta \cite{fu2010odd}. Existing band structure calculations have found that this is indeed the case in $\beta$-PdBi$_2$ (See Fig. 2a) \cite{sakano2015topologically}. Thus, K-doped $\beta$-PdBi$_2$ is likely to be a topological superconductor, and could undergo an unprecedented 3D topological superconducting phase transition between a trivial and a topological superconductor. If there is an intermediate magnetic phase, the TSC-magnetism critical point would be a condensed matter realization of supersymmetry \cite{grover2014emergent}. In addition, the transport properties across the across the topological quantum phase transition can be useful for technological applications.

$\beta$-PdBi$_2$ has a layered, tetragonal crystal structure, belonging to the I4/mmm space group. Band structure calculations \cite{shein2013electronic, sakano2015topologically} show that the compound is a multi-band metal, with two hole-like and two electron-like pockets formed by Bi-6p and Pd-4d bands at the Fermi level ($E_f$). The Bi-6p and Pd-4d bands are already inverted without spin-orbit coupling (SOC) \cite{shein2013electronic, sakano2015topologically}, and the role of SOC is to drive the opening of a bulk band gap below $E_f$, i.e., to create a metallic band structure that can be transformed into that of an insulator by smoothly deforming the bands and lowering $E_f$. In addition, (spin-) angle-resolved photoemission spectroscopy (spin-ARPES) and quasiparticle interference imaging have revealed the presence of spin-polarized topological surface states around $E_f$ \cite{sakano2015topologically, iwaya2017full}. Intrinsic superconductivity robust to different dopants \cite{dopant, zhao2015chemical} and a relatively high T$_c$ compared to other systems make this material an attractive candidate for realizing topological superconductivity. Until now, however, transport properties of the topological surface states in single crystals of $\beta$-PdBi$_2$ have been difficult to measure due to the coexistence of trivial bands with non-trivial surface bands around $E_f$.

\section{\label{sec:level1} NORMAL STATE TRANSPORT PROPERTIES}

\begin{figure*}[]
\centering
\includegraphics[width=0.7\textwidth,,height=14cm]{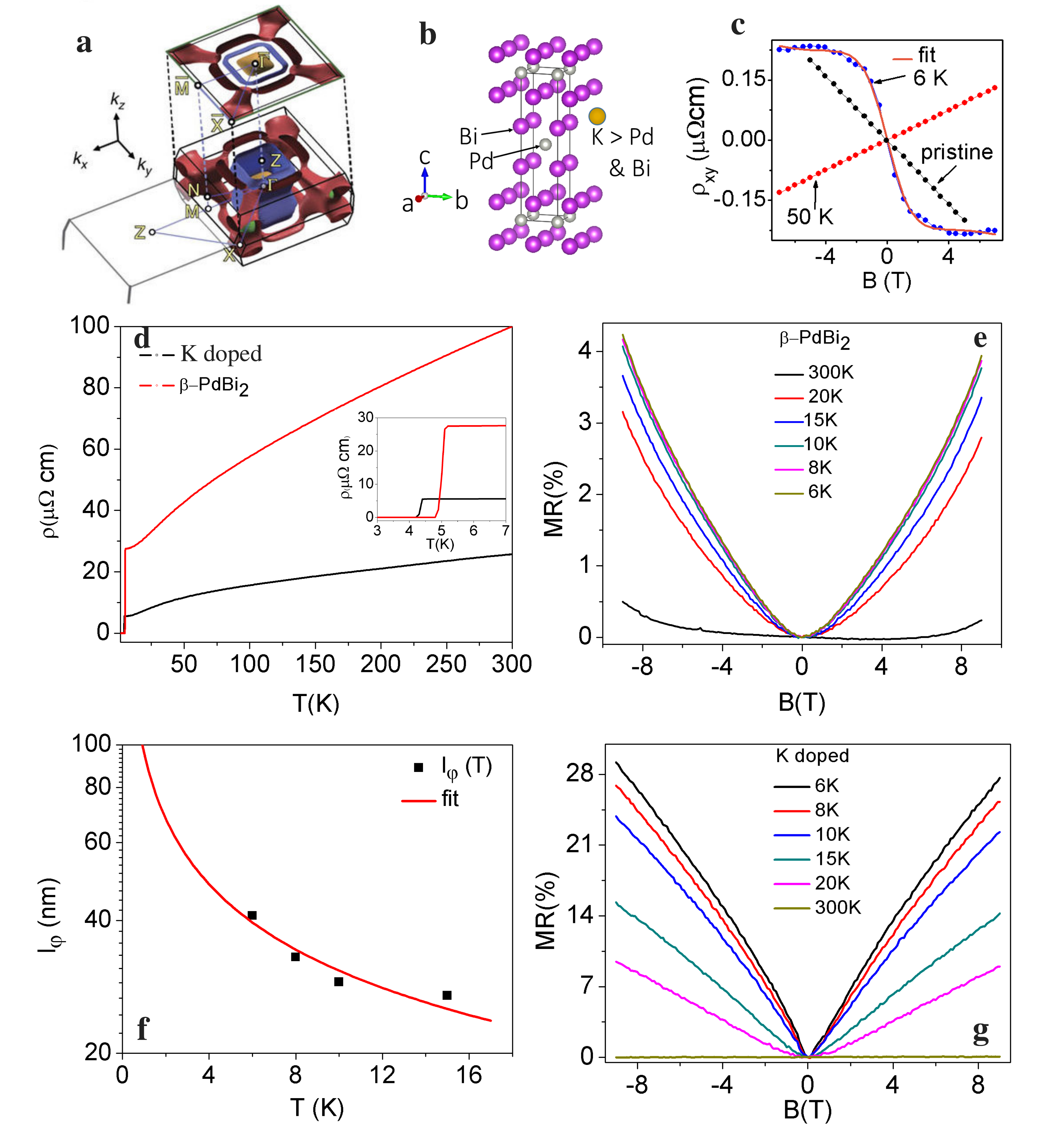}
\caption{ \textbf{Transport properties of K doped and pristine $\beta$-PdBi$_{2}$ above $T_c$ } (a) Fermi surfaces of $\beta$-PdBi$_{2}$ as calculated in ref. \cite{sakano2015topologically} showing that the Fermi surfaces enclosed an odd number of time-reversal invariant momentum points. (b) Crystal structure of $\beta$-PdBi$_{2}$. Note that K atom is bigger than Pd and Bi atoms. (c) Symmetrized Hall resistivity of K doped $\beta$-PdBi$_{2}$ at 6 K showing the non-linear behavior characteristic of two channel systems (see Supplementary Information S2 Ref. \cite{Supplemental}). The pristine $\beta$-PdBi$_{2}$ data in c is from \cite{zhao2015chemical}. (d) Electrical resistivity along the a-b plane of pristine and K doped $\beta$-PdBi$_{2}$ as a function of temperature (T).  (e) LMR in $\beta$-PdBi$_{2}$: There is no clear sign of WAL as the LMR scales with B$^2$. (g) LMR in K doped $\beta$-PdBi$_{2}$: WAL can be seen clearly from 15 K till the lowest temperature (6 K) in our experiment. At higher temperatures, the phase coherence length becomes shorter and quantum interference effects are lost, resulting in an LMR that scales with $B^2$ at 20 K and above in the low field range, and (f) is temperature dependence of the phase coherence length.} 
\end{figure*}

We begin by briefly studying the transport properties above the superconducting transition temperature of pristine ($T_c$ = 5.3 K) and potassium doped (0.3 \%) $\beta$-PdBi$_2$ ($T_c$ = 4.4 K). In Fig. 2, we explore the temperature and magnetic field dependence of the longitudinal and Hall resistivity. Potassium can be an ambipolar dopant in this system. It can either enter interstitial sites and serve as an electron donor, or replace the elements and become an acceptor. We find evidence in the Hall resistance curves (Fig. 2c) that K acts as an acceptor. At above 20 K, it changes transport from electron- to hole-dominated upon doping, while the low temperature data exhibits a non-linearity characteristic of two-channel conduction -- a highly mobility (4280 cm$^2$/Vs) electronic channel and relatively lower mobility (32 cm$^2$/Vs) hole channel. (See Supplementary Information S2 Ref. \cite{Supplemental}). Following Ref. \cite{ando2013topological}, we attribute the high mobility to surface Dirac electrons and the low mobility to holes in the bulk. At higher temperatures, the quantum effect of the surface Dirac electrons is diminished and the bulk carriers dominate transport in the doped system, resulting in a classical, hole-dominated linear Hall resistance.

The normal state LMR $\rho_{xx}(B)$, of doped $\beta$-PdBi$_2$ (Fig. 2g) is compared with the pristine sample (Fig. 2e). The LMR of $\beta$-PdBi$_2$ lacks clear evidence of WAL as it scales with $B^2$ in the lower B field range. The doped crystal, in contrast, shows a sharp magnetoresistance cusp characteristic of WAL. WAL is a striking manifestation of the $\pi$ Berry phase of spin-momentum locked states on the surface of a TI, which protects them against localization by non-magnetic impurities \cite{ando2013topological}. This effect is quickly suppressed in the presence of magnetic field which breaks the time-reversal symmetry and results in a sharp increase in resistance, reflected by the cusp in the magnetoresistivity. The conductivity for small $B$ can be described by Hikami-Larkin-Nagaoka (HLN) formula \cite{hikami1980spin}. At 6 K, the phase coherent length $l_\phi$ = 41 nm, and it rapidly decreases with temperature. For 2D transport, the temperature dependence is expected to follow  \cite{altshuler1982effects}  $l_\phi(T)$ $\propto$ $T^{-0.5}$;  in Fig. 2f the temperature dependence can be fitted with $l_\phi(T) \propto T^{-0.49}$, suggesting that the WAL occurs in the 2D surface states.

The above discussion provides evidence that doping depletes the trivial bands and  exposes the spin-polarized topologically non-trivial surface bands which have been already measured by spin-resolved ARPES \cite{sakano2015topologically, iwaya2017full}, thus unveiling its topological nature in the normal phase. We are now well-positioned to explore the effects of topological state on superconductivity, especially, on Cooper pairing in the bulk and its interplay with the band structure to produce non-trivial topology in the superconducting phase.

\section{\label{sec:level1} SUPERCONDUCTING TRANSPORT PROPERTIES}
\subsubsection{Point-Contact Spectroscopy}
To investigate the surface of the superconductor, we performed 'soft' point-contact spectroscopy \cite{ daghero2010probing} (See Supplementary Information S3 Ref. \cite{Supplemental} ) on K-doped $\beta$-PdBi$_2$ cooled down to 300 mK, studying the magnetic field and temperature dependence of the differential conductance, d$I$/d$V$. We present the magnetic field dependence of d$I$/d$V$ with current along the $ab$ plane and the magnetic field along the $c$ axis in Fig. 3a.  z is representative of the barrier strength: z = 0, Andreev spectroscopy; while z = $\infty$ (z $\sim$ 5 in experiments),  tunneling spectroscopy. Here is z = 0.4. We note that the spectrum at zero magnetic field looks remarkably different from the rest. In particular, in Fig. 3b (where we have normalized d$I$/d$V$ to 1 as V $\rightarrow$ $\infty$), we see conductance dips at $\pm$ 1 meV and peaks exceeding the value predicted by the Blonder-Tinkham-Klapwijk (BTK) formalism of Andreev reflection \cite{blonder1982transition} for a conventional SC-insulator-normal metal interface (with z = 0.4). In fact, the peaks at 0.3 K exceed the theoretical maximum value of 2 (required by Andreev process) for a gapped superconductor, and might thus be indicative of gapless states.

Let us recall the known causes of conductance dip in PCS d$I$/d$V$ spectra. (i) \textit{Critical current or heating effect}\textemdash dips at positions larger than the superconducting energy gap are often found in the spectrum when the contacts on the sample are in the thermal limit \cite{sheet2004role}. At the critical current the superconductor turns into a normal metal, and when measurements are carried out in the thermal limit, the resistance of the bulk sample is measured in the d$I$/d$V$ spectrum. Since the critical current required to limit superconductivity reduces with increasing magnetic field and temperature, these dips are found to occur at position of decreasing lower bias voltage. In our experiments, the dip position does not reduce with increase in temperature and magnetic field (in fact it was only observed at zero magnetic field), so the critical current effect is ruled out (See Supplementary Information S4 Ref. \cite{Supplemental}). (ii)  \textit{ 1D, 2D and 3D TSC} \textemdash topological superconductors feature dips at $\pm \Delta$. A simple physical explanation for this is the transfer of spectra weight from the states near the gap to make up for the in-gap states.  In addition to the dips, 1D and 2D TSCs feature zero-bias conductance peak (ZBCP) due to Andreev bound state (ABS), while TRI 3D TSCs do not \cite{ mourik2012signatures, kashiwaya2014tunneling, sasaki2011topological, yamakage2012theory, sato2017topological}.

For a finite potential barrier between the contact and an ideal 3D topological superconductor, dI/dV $\propto$ surface density of states, and differential conductance spectrum should produce a double peak structure \cite{sato2017topological, sasaki2011topological, yamakage2012theory} \textemdash that is, ZBCP is \textit{not} expected for fully gapped TRI 3D topological superconductors. It should be pointing out, however, that the tunneling conductance can feature a ZBCP due to various effects. In studies for superconducting 3D TI, remnant Dirac fermions from the normal state are found to modify in the superconducting state in two ways: one, it enhances the pair potential resulting in a larger gap for the surface superconductivity \cite{mizushima2014dirac}. Two, if the Dirac surface state is well separated from the bulk, it can twist the surface Majorana cone, resulting in the ZBCP \cite{yamakage2012theory}. Furthermore, if the bulk superconductivity is not fully-gapped as is the case for Cu$_x$Bi$_2$Se$_3$, for example, the tunneling conductance features a ZBCP \cite{sasaki2011topological}. Otherwise in the ideal case, the differential conductance features a double peak. Here, the presence of dI/dV$_{norm}$ $>$ 2 (for z = 0.4) and the non-trivial conductance dips at zero magnetic field are in good agreement with the theoretical prediction for 3D TRI TSCs.

\begin{figure*}[t!]
\centering
  \includegraphics[width=0.7\textwidth,height=12cm]{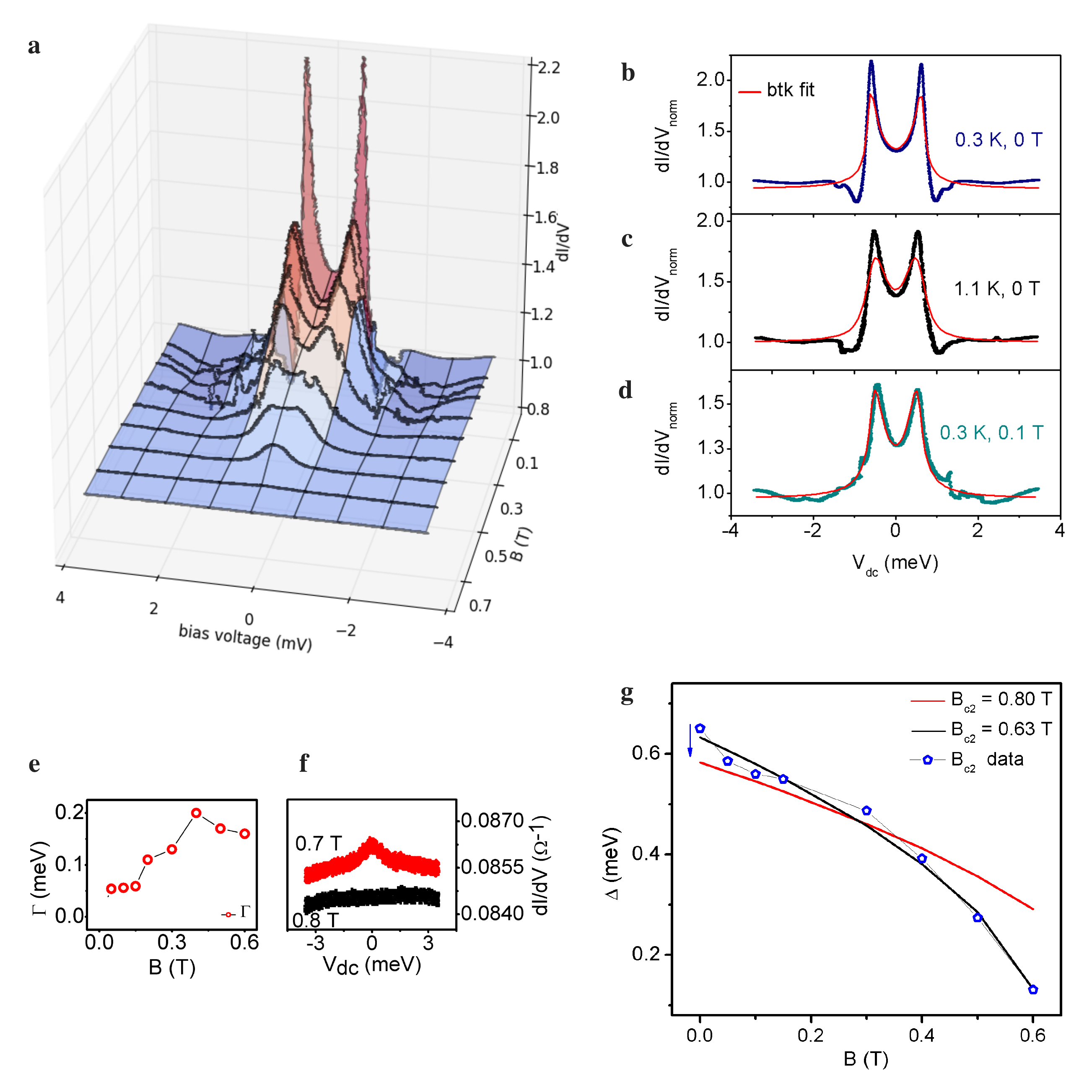}
  \caption{ \textbf{Point-contact spectroscopy.} (a) Magnetic field dependence of the d$I$/d$V$ vs bias voltage for K doped $\beta$-PdBi$_{2}$ at 0.3 K. (b, c) BTK fitting of the d$I$/d$V$ spectrum at 0 T for 0.3 K and 1.1 K. The fit is poor. In comparison, the fit is good at 0.1T in (d). (e) Field evolution of the quasi-particle lifetime broadening parameter $\Gamma$.  (f) close up view of d$I$/d$V$ vs bias voltage at 0.7 T and 0.8 T. (g) Attempts to fit the gap with the BCS magnetic field dependence equation.}
\end{figure*}

\begin{figure}[t!]
  \centering
  \includegraphics[width=0.5\textwidth,height=8cm]{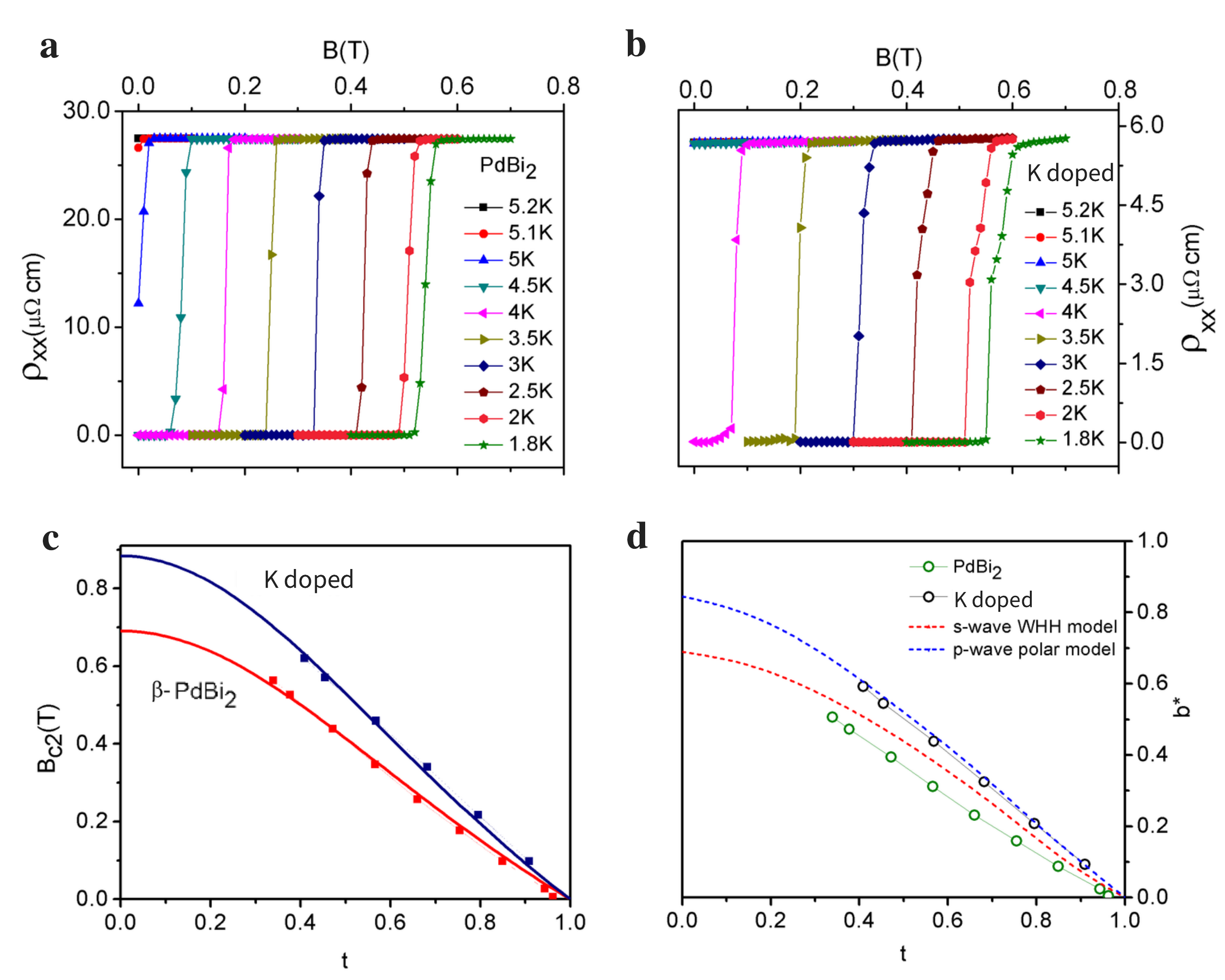}
  \caption{\textbf{Upper-critical field analysis.} (a) and (b), Variation of the upper critical field $B_{c2}$ as a function of temperature in pristine and K doped $\beta$-PdBi$_{2}$. (c)  the Ginzburg-Landau fit is shown in red and blue for pristine and K doped $\beta$-PdBi$_2$ respectively. (d) Plot of the reduced upper critical field, $b^* = B_{c2}/|dB_{c2}/dt|_{t = 1}$ as a function of the reduced temperature $t = T/T_c$. The red dash is the upper limit for s-wave superconductivity according to the WHH model. A conventional superconductor with finite SOC and Maki parameter will be below the universal WHH model curve. K doped $\beta$-PdBi$_2$ lies above the upper-limit of WHH model and closer to the polar p-wave model, pointing to K doped $\beta$-PdBi$_2$ as an odd-parity superconductor.}
\end{figure}

\begin{figure}[]
\begin{center}
\includegraphics[width=0.5\textwidth,height=12cm]{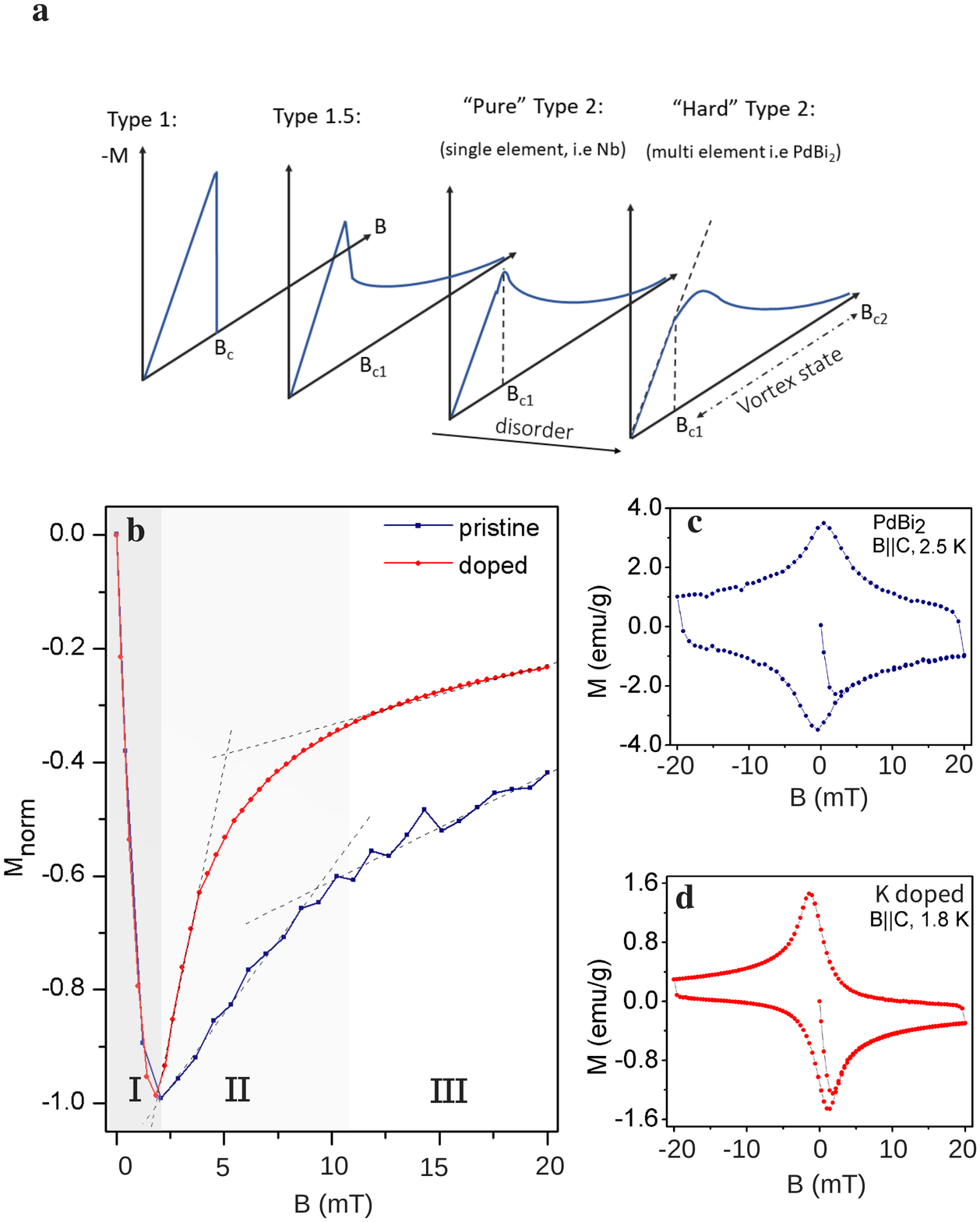}
\end{center}
\caption{\textbf{Spin-triplet Vortex State.} (a) $B_{c1}<B<B_{c2}$ is the vortex state: flux vortices starts to penetrate the superconductor at $B_{C1}$ and form a vortex lattice leads to irreversibility of the magnetization loop due to flux pinning effects. At the irreversibility field, the vortex lattice melts, restoring the reversibility. Doping increases the the number of pinning sites and consequently, irrersibility. However, a spin-triplet superconductor can demonstrate  the opposite behavior, dubbed `type 1.5'. (b) Comparison of the normalized magnetization along the $B||c$ plane of both the undoped and doped sample, at 2.5 K and 1.8 K respectively. Beyond $B_{c1}$ of K doped $\beta$-PdBi$_{2}$ the magnetization shows an anomalous increase in magnitude which is at odds with the conventional type-II "alloy" superconductors but consistent with the expectation for a spin-triplet pairing. (c)-(d) magnetization for the pristine and doped sample respectively.}\label{lines}
\end{figure}

Looking beyond the presence or absence of the ZBCP as a precise evidence of non-trivial superconductivity, we study the effect of a time-reversal breaking perturbation, that is, magnetic field on the surface states. Analogous to the time-reversed Dirac fermions on the surface of a TI which are protected from backscattering, the superconducting state is expected host helical pairs of Majorana fermions which are robust against non-magnetic disturbances \cite{xu2014momentum, qi2009time}. This physics is captured here: applying a magnetic field breaks time-reversal symmetry\textemdash and protection from scattering is lifted. The helical surface states is destroyed and the underlying superconductivity described by the usual BTK-like spectrum is uncovered. We see in Fig. 3b-d that the spectrum under 0.1 T at 0.3 K fits the BTK model for conventional superconductivity \cite{blonder1982transition, dynes1978direct} while that under zero magnetic field does not. Note that 0.1 T $\ll B_{c2}\sim$ 0.8 T (as estimated below), so modifications to d$I$/d$V$ data due to vortices and due to suppression of the bulk gap are expected to be negligible.

We extracted the superconducting gap at different magnetic fields using the BTK formalism and attempted to fit its evolution with the prediction for a conventional Bardeen-Cooper-Schrieffer (BCS) superconductor: $\Delta(B) = \Delta_0 (1 - B/B_{c2} )^{1/2}$. Experimentally, $B_{c2}$ is found to be somewhere between 0.7 and 0.8 T according to the B field dependence of Andreev reflection as shown in Fig. 3f. However, the gap could not be described by BCS using 0.7 T $< B_{c2} <$ 0.8 T ; the misfit for 0.8 T is shown in 3g. By making both the $\Delta_0$ and $B_{c2}$ free parameters, we got the best fit with 0.63 T. Clearly, as shown in Fig. 3f the crystal is still superconducting up till at least 0.7 T. This proposes that the bulk superconducting state might not be entirely described by BCS theory. Odd-parity, unconventional bulk superconductivity is a necessary condition for 3D topological superconductivity -- in which the just demonstrated helical surface states are a signature. The next step, therefore, is to study the bulk superconductivity.

\subsubsection{Upper-Critical Field Limiting Effect}
To better understand the nature of the bulk superconducting state from which the surface state arises, we proceed to study the upper critical field limiting effect. In Fig. 4, the upper critical field $B_{c2}$ at different temperatures below $T_c$ is plotted and extrapolated to T = 0 by using the Ginzburg-Landau form, $B_{c2}(t) = B_{c2}(0)(1- t^2)/(1 + t^2)$,  where $t = T/T_c$. $B_{c2}(0)$ is found to be 0.89 T for K doped crystal. This value is clearly higher than the value of 0.63 T obtained by fitting the d$I$/d$V$ data to the BTK formula for a conventional BCS SC, confirming that the superconductivity is unconventional.

It is remarkable that doped $\beta$-PdBi$_2$ has lower $T_c$ but higher $B_{c2}(0)$, as this is suggestive of spin-triplet pairing. Indeed, the mean free path $l$ is greater than the coherent length $\xi$, and the WHH and Pauli limiting effects are absent(See supplementary Information S5 Ref. \cite{Supplemental}).  We can gain more insight by comparing the $B_{c2}(T)$ data with the well known theoretical model for s-wave and polar p-wave \cite{werthamer1966temperature, scharnberg1980p}. Fig. 4 is the plot of the reduced upper critical field, $b^* = B_{c2}/|dB_{c2}/dt|_{t = 1}$ versus the reduced temperature $t = T/T_c$ compared to the theoretical models for s-wave and polar p-wave SCs.  For the doped crystal, we note that the experimental data fits better to the p-wave than to the s-wave model. The pristine $\beta$-PdBi$_2$ in comparison lies below the upper limit of the WHH s-wave theoretical prediction. The s-wave WHH model presented in Fig. 4 is the upper limit, derived for $\alpha$ = $\lambda_{so}$ = 0, where $\alpha$ and $\lambda_{so}$ are the Maki parameter \cite{maki1966effect} and the spin-orbit strength, respectively; non-zero $\alpha$ and $\lambda_{so}$ moves the curves lower. With $\alpha$ = $0.53 |dB_{c2}/dT|_{T_c}$ = 0.11 and a finite $\lambda_{so}$ \cite{shein2013electronic} in $\beta$-PdBi$_2$, the pristine crystal can be well described by the WHH model. This is in contrast to the doped crystal where b$^*$ lies above the s-wave WHH upper limit. For further evidence of unconventional superconductivity, we carried out magnetization experiments to study the vortex state.

\subsubsection{Magnetization: Vortex State}
The magnetization for a spin-triplet superconductor, as demonstrated for Cu$_x$Bi$_2$Se$_3$  \cite{das2011spin}, exhibits the so-called Type 1.5 like behavior (see Fig. 5). Consider the effect of the magnetic field induced by the persistent vortex current on the spins of the spin-triplet pairs. The induced magnetic field polarizes the Cooper pairs and an additional spin magnetization arises. The total magnetic flux in the vortex now consists of the current and spin magnetization contributions \textemdash and is quantized. The quantization of magnetic flux causes current inversion in parts of the vortex, favoring their formation just above $B_{c1}$ by driving an attractive interaction between the vortices. This explains the anomalous increase of the magnetization past the $B_{c1}$ threshold and the low irreversibility often observed in the magnetization vs magnetic field loop of Cu$_x$Bi$_2$Se$_3$. We also observe the sharp increase in magnetization beyond $B_{c1}$ and subsequent low irreversibility in doped PdBi$_2$ (which is absent in the pristine sample) and propose that this effect can be explained by spin-triplet pairing.

In the Fig. 5, we compare the magnetization of the pristine crystal with the doped one. Below $B_{c1}$, labeled region I, diamagnetization occurs at the same rate for both samples; above $B_{c1}$ the effect of current inversion is observed in the doped crystal. As the vortices becomes attractive, flux vortices are forced in the doped sample and the magnetization increases sharply in region II and continues to do so till the beginning of region III. In the third region, the magnetization rate saturates because the repulsive force between vortices due to high density counterbalances the attractive force effect. Also the vortex-lattice state melts into vortex-liquid state at a faster rate; thus the low irreversibility in spite of the doping.

\section{\label{sec:level1} DISCUSSION}
Having studied various bulk and surface transport properties in the normal and superconducting phases, we address the question of \emph{why} the doped system behaves so differently from the undoped one.

\textit{Effects on surface states in the normal state}: Firstly, the nontrivial surface states around E$_f$ in the normal phase coexist with trivial surface and bulk bands \cite{sakano2015topologically}. While a direct probe of the band structure such as spin-ARPES can detect the topological surface states, it is difficult to probe them in a transport experiment. One route for reducing the bulk contributions is using thin films. Alternately, appropriate dopants can deplete the trivial bands, uncovering the topological ones, and we find that K doping indeed accomplishes this goal. 

\textit{Effects on bulk superconductivity}: Secondly, we recall that sufficient conditions for topological superconductivity in a 3D TRI material are that the normal state Fermi surfaces enclose odd number of TRIM \emph{and} the fullu-gapped bulk superconductivity pairing be odd under inversion. In pristine $\beta$-PdBi$_2$ only the former condition met. Bulk superconductivity is s-wave \cite{2016single,herrera2015magnetic,che2016absence}, so topological superconductivity is not expected (see lower-left corner of Fig. 1); instead, one gets a Fu-Kane-like superconductor with surface Majorana zero modes in vortex cores \cite{fu2008superconducting, lv2017experimental}, which are distinct from 2D helical Majorana surface states. On doping with K, we find that the c-lattice parameter increases (see Supplementary Information S1 Ref. \cite{Supplemental}) and transport changes from electron to hole dominated, indicating that larger K$^+$ ions have replaced the smaller Pd and Bi ions. This would lead to local inversion symmetry breaking without breaking the centrosymmetry of the bulk crystal, which is enough to induce spin polarizations in the bulk of layered, centrosymmetric systems \cite{zhang2014hidden, riley2014direct}. It is now well known that in the presence of  spin-orbit-coupling, electron-phonon interactions can give rise to even- as well as odd-parity pairing \cite{brydon2014odd, kozii2015odd, wang2016topological}. The s-wave, even-parity state invariably onsets at a higher $T_c$ \cite{brydon2014odd}, driving the odd-parity state to T = 0. Suppressing the s-wave channel, for example, by adding minute local spin-polarization could promote the odd-parity channel. In centrosymmetric Bi$_2$Se$_3$, superconductivity is induced by intercalating A = Cu, Nb, Sr into the non-superconducting parent compound. Recent spin, transport and thermodynamics experimental studies on  A$_x$Bi$_2$Se$_3$ have shown the evidence of 2-fold pairing symmetry consistent with odd-parity, nematic superconductivity  as opposed to the 6-fold symmetry of the hexagonal Bi$_2$Se$_3$ structure \cite{yonezawa2017thermodynamic, pan2016rotational, asaba2017rotational}.

We emphasize that the bulk superconductivity intrinsic to $\beta$-PdBi$_2$ opens the possibility of a topological phase transition in the superconducting state from a trivial to a topological superconductor, which prior to this work has not been realized because undoped Bi$_2$Se$_3$ does not host intrinsic superconductivity. In case the topological superconductor first transitions into a magnetic phase, the accompanying critical point will present a unique laboratory realization of emergent supersymmetry \cite{grover2014emergent}. Thus, $\beta$-PdBi$_2$ with intrinsic superconductivity and a centrosymmetric layered structure presents a unique and robust platform for exploring various aspects of 2D helical Majorana fermions, just as 3D topological insulators enabled the study of 2D helical electrons on their surface. 

\section{\label{sec:level1} MATERIALS AND METHODS}
$\textbf{Crystal growth and characterization.} $ Single crystals of $\beta$-PdBi$_2$ were successfully synthesized by solid-state melt method. Stoichiometric amounts of Pd grains and Bi powder, mixed at a molar ratio of 1:2, were sealed in an evacuated quartz tube and ran through the thermal profile as follows: the mixture was heated up to 900$^o$C in 4 hours and left to melt and mix for 20 hours. It was then cooled down at 3$^o$C/ hour to allow for crystal formation and swiftly quenched in iced water at 480$^o$C. The quenching is best done above 450$^o$C to prevent the formation of the $\alpha$-phase. The potassium doped sample was synthesized by nominally replacing Bi with the dopant element. The composition was characterized by wavelength dispersive spectroscopy; chemical analysis puts the actual potassium content much lower than the starting composition. To confirm good crystallization, we carried out x-ray diffraction experiments-- all the peaks were index and no traces of the $\alpha$-phase was detected. Furthermore, the doped crystal was checked for single crystalline domain orientation by Laue back-reflection diffraction method. Our Laue analysis on the pattern of dots indicates the formation of single crystalline domain orientation.

$\textbf{Transport measurements.}$ The samples were first cut into rectangular shape, then cleaved into the desired thickness by gently peeling of the layered crystal layers by layers using scotch tape. The freshly cleaved surface which is mirror-like and shiny is then gold-plated at the contact points to achieve better ohmic contact. The four-probe technique was used for the longitudinal resistance, $R_{xx}$, and the Hall resistance, $R_{xy}$, was acquired by the standard method. The magneto-resistance and magnetization measurements were carried out using Quantum Design Inc.'s Physical Properties Measurement System (PPMS) and Magnetic Property Measurement System (MPMS) respectively. The point-contact spectroscopy was performed in an He-3 refrigerator.
 
\begin{acknowledgements} 

The authors wish to acknowledge R. Forrest, K. Dahal, Y. Lyu, S. Huyan and U. Saparamadu for technical assistance. We thank W.-P. Su, C. Ting, A. Guloy, D. Stokes, Lei Hao, Hai Li and R. Du for helpful discussions. The work in Houston is supported in part by the State of Texas through the Texas Center for Superconductivity at University of Houston (AK and JHM); and the U.S. Air Force Office of Scientific Research Grant No. FA9550-15-1-0236, and the T.L.L. Temple Foundation, the John J. and Rebecca Moores Endowment (AK); and by the Division of Research, Department of Physics and the College of Natural Sciences and Mathematics at the University of Houston (PH). TL is supported by NSF grant number DMR-1508644.

\end{acknowledgements}

\putbib

\end{bibunit}

\onecolumngrid
\begin{center}
  \textbf{\large Supplementary material for: Transport evidence for three-dimensional topological superconductivity in doped $\beta$-PdBi$_2$}\\[.2cm]
  Ayo Kolapo,$^{1,*}$ Tingxin Li,$^{2}$, Pavan Hosur$^{1,*}$ and John H. Miller$^1$ \\[.1cm]
  {\itshape ${}^1$Texas Center for Superconductivity and Department of Physics, University of Houston, 3201 Cullen Boulevard, Houston, Texas 77204, USA,\\ Department of Physics and Astronomy, Rice University, Houston, Texas 77251, USA\\}
  ${}^*$email: aykolapo@uh.edu, phosur@uh.edu\\
(Dated: \today)\\[1cm]
\end{center}

\setcounter{equation}{0}
\setcounter{figure}{0}
\setcounter{table}{0}
\setcounter{page}{1}
\renewcommand{\theequation}{S\arabic{equation}}
\renewcommand{\thefigure}{S\arabic{figure}}

\begin{bibunit}

Here we present the x-ray diffraction (XRD), Hall resistivity analysis, 'soft' point-contact spectroscopy experimental methods, and analysis of WHH orbital and Pauli limiting effect on potassium doped $\beta$-PdBi$_{2}$. 

\section*{S1. X-ray Diffraction}
Single crystalline samples of pristine and 0.3 \% K doped $\beta$-PdBi$_2$ were characterized by XRD using the Rigaku DMAX IIIB diffractometer at room temperature with a Cu K ($\lambda$ $=$ 0.154056 nm) radiation source. The $hkl$ ($00l$) reflection patterns for undoped and K doped $\beta$-PdBi$_2$ are shown in FIG. S1. The inset shows the 2$\theta$ shift in the doped crystal revealing a 0.3 \% increase in the c lattice parameter.

\begin{figure}[!htbp]
\centering
\includegraphics[width=13cm]{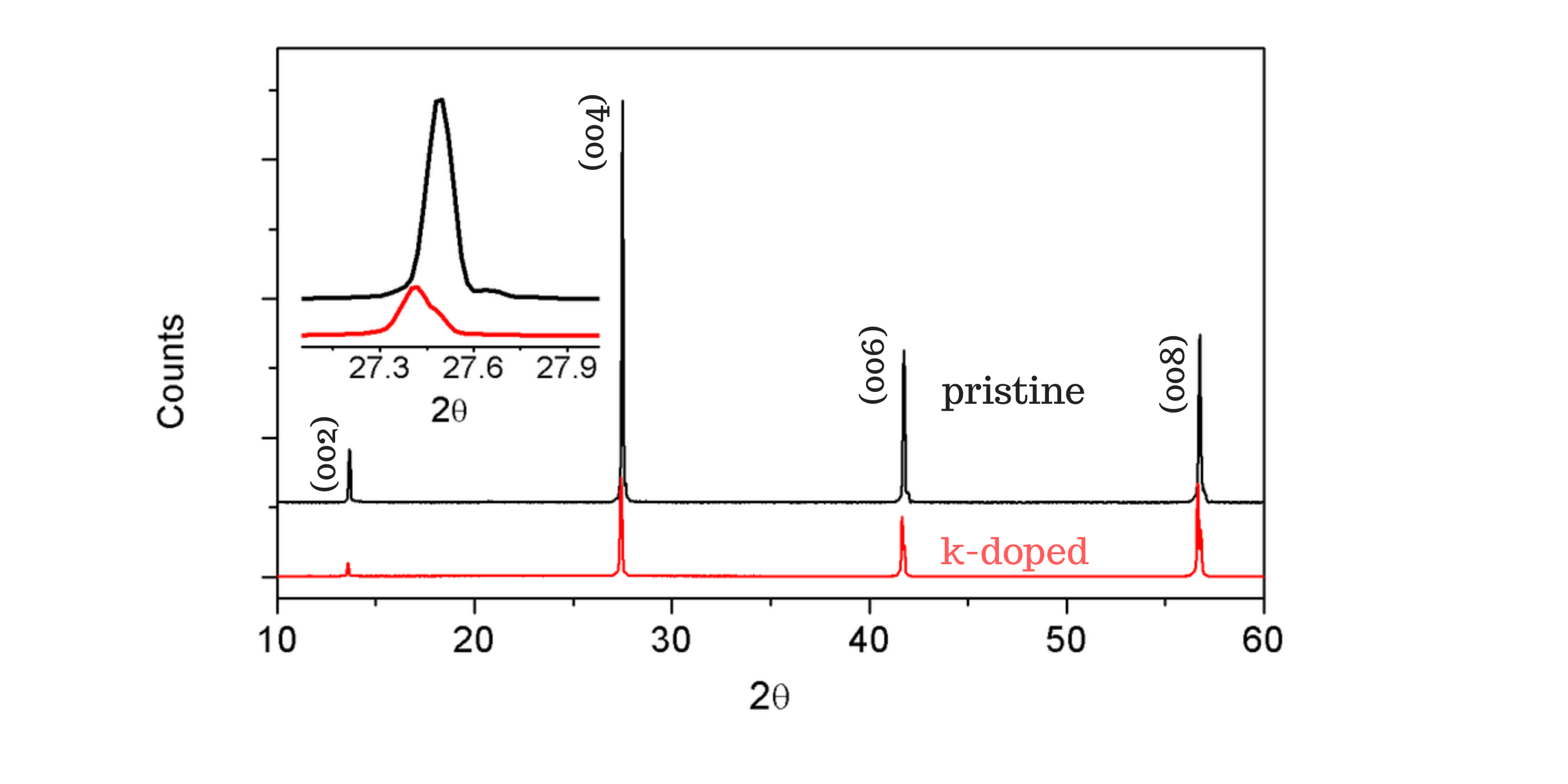}
\caption{Single crystal x-ray diffraction for pristine (black line) and K-doped $\beta$-PdBi$_{2}$ crystal (red line)}
\end{figure}

\pagebreak

\section*{S2. Hall Resistivity: Two-Channel Analysis}

The Hall resistivity data show non-linear behavior below 20 K, while it display linear behavior from above 20 K up to room temperature. In topological materials, this non-linearity is due to parallel contribution from the surface and bulk states \cite{ando2013topological, qu2010quantum}. 

Both the surface sheet and the bulk contributions to the Hall conductivity can be analyzed using \cite{ashcroft1976nd}:

\begin{equation}
\rho_{xy} = \frac{ ( R_s \rho^2_b + R_b \rho^2_s)B +  R_sR_b (R_s + R_b) B^3}
{ (\rho_b + \rho_s)^2 + (R_s + R_b)^2 B^2 } 
\end{equation}
            
where $\rho_s $ and $\rho_b$ is the surface sheet and bulk resistivity respectively; and $R_s$ and $R_b$ is the surface sheet and bulk Hall coefficient respectively. Here $R_b= \frac{1}{en_b}$, $R_s = \frac{t}{en_s}$, and $\rho_s = \frac {1}{\sigma_s} = \frac {t}{G_s} $.                                        

Using the value the longitudinal resistivity at zero field as a constraint, that is, $ \rho_{xx}^{-1}$ = $\rho_{s} ^{-1} + \rho_{b} ^{-1}$, we fit the Hall resistivity curves as shown in Fig. S2. It can be fitted well with the 2-channel model. 

From the fitting value of $R_s$, we estimate $n_s$ = $- 7.7 \times 10^{16}$ \ cm$^{-2}$ for the surface carrier density and $n_b$ = $ 3.4 \times 10^{22}$ \ cm$^{-3}$ for the bulk. From these derived fitting parameters of $\rho_{s}$ and $\rho_{b}$, the percentage of surface contribution is calculated as:
\[ \frac{G_s}{G_s + \sigma_b t } = 0.06\]

estimating that $\sim$ 6 \% of the total conductance is due to the surface states at 6 K. 
The mobility for the surface electron is $\mu_s = \frac{R_s}{\rho_s}$ = 4280 \ cm$^2/Vs$ which is 134 times larger than the mobility of the bulk carriers $\mu_b = \frac{R_b}{\rho_b} =$ 32 \ cm$^2/Vs$. With this high mobility, the surface states dominate the transport at low temperatures.

\begin{figure}[b!]
\centering
\includegraphics[width=0.5\textwidth,,height=6.5cm]{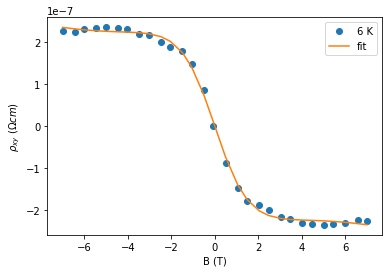}
\caption{Symmetrized Hall resistivity of K doped $\beta$-PdBi$_{2}$ at 6 K showing the non-linear behavior characteristic of two channel systems. } 
\end{figure}

\pagebreak

\section*{S3. Point-contact Spectroscopy}
The point-contact spectroscopy set up is shown in Fig. S3. Here, ballistic transport is achieved through several channels of nanometer sized silver particles contained in the silver epoxy \cite{daghero2010probing, sasaki2011topological}. This method is in contrast to the hard metallic etched tip used in Scanning tunneling spectroscopy (STS), hence, the 'soft' appellation. The d$I$/d$V$ measurements were obtained by superimposing a constant 5 $\mu$A AC current with a sweeping DC current. The corresponding AC voltage can be picked up with great accuracy by the lock-in amplifier. The experiments were then carried out in a He-3 refrigerator.

\begin{figure}[h]
\begin{center}
\includegraphics[width=8cm]{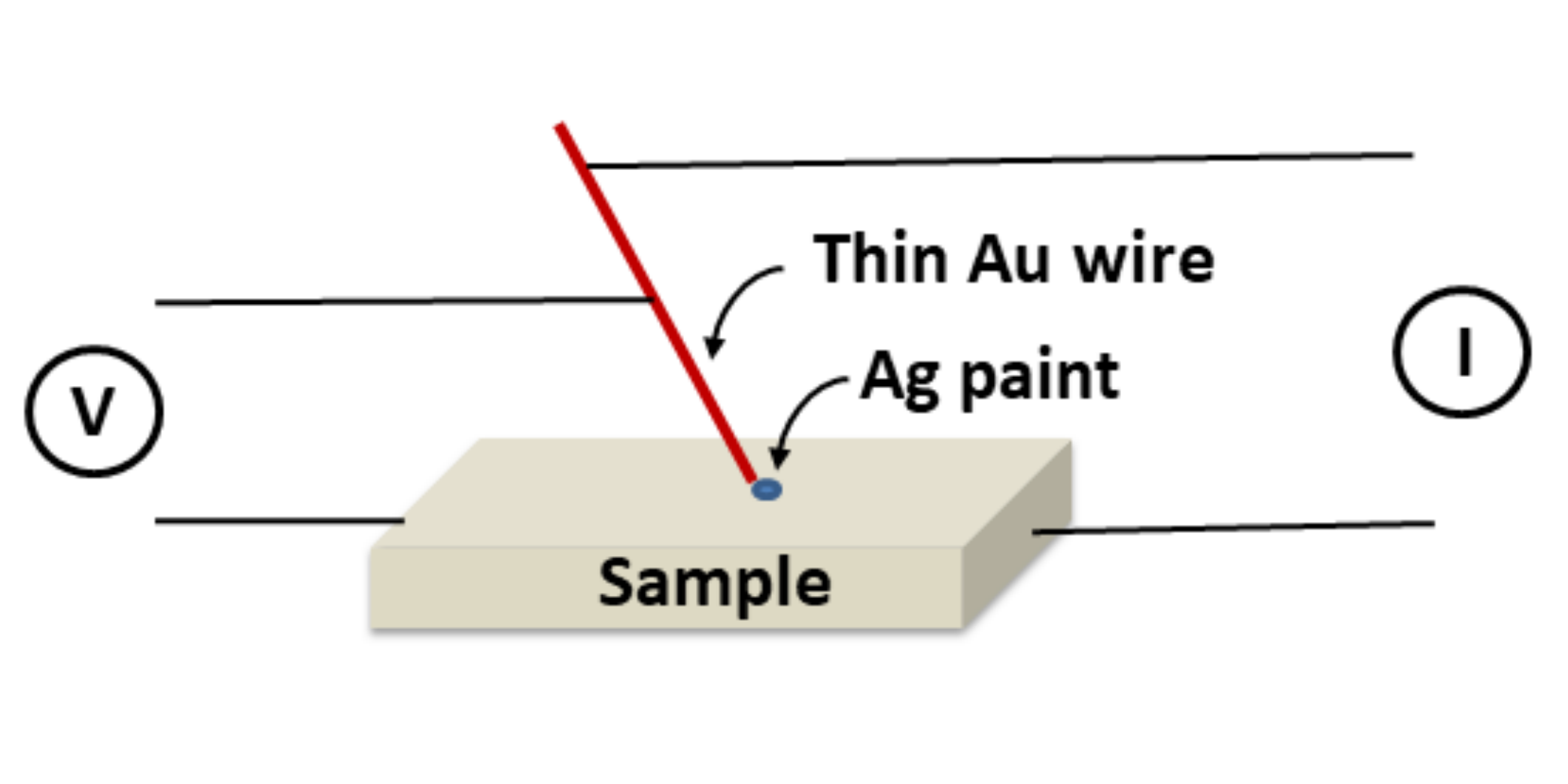}
\end{center}
\caption{'soft' point-contact spectroscopy set up. The current is past through the the thin Au wire to the sample through a 30 $\mu$m tiny drop of Ag nano-particle epoxy paint.}
\end{figure}

\vspace{20mm}

\section*{S4. Differential Conductance}
Here we comment and highlight on the feature in the d$I$/d$V$ spectrum presented in the main text. 

The BTK formalism for the Andreev reflection can be written as:

\begin{equation}
\frac{dI}{dV}_{Norm} \propto \int ^{\infty} _{-\infty} dE \frac{df(E+eV}{d(eV)} [1 + A(E) - B(E)]
\end{equation}

where $ { \frac{dI}{dV} }_{Norm}$  =  $\frac{{\frac{dI}{dV}}_{NS}}{{\frac{dI}{dV}}_N} $, and $A(E)$ is the probability amplitude for Andreev reflection, and $B(E)$ is the amplitude for specular reflection. At $T$ = 0 and in the limit z $\rightarrow$ 0, then for a particle with $E < \Delta$ undergoing a complete Andreev reflection, we have $\frac{dI}{dV}$ = 2, because B(E) = 0. When $T$ and $B(E)$ is finite,  $\frac{dI}{dV} <$ 2.

In our experiments, the spectroscopy  is consistent with the BTK formalism when the magnetic field is turned (that is, when the magnetic field localizes the helical surface states). In the absence of the magnetic, we see the normalized differential conductance suddenly rising to $>$ 2. We also see that the conductance peak around $\pm \Delta$ is not due to critical current. In this paper, we have shown that K-doped $\beta$-PdBi$_{2}$ satisfies the necessary condition for bulk topological superconductivity, therefore an helical surface surface is guaranteed to exist (whether its detected in transport measurements or not). Then, it is reasonable to suggest that the anomalous zero magnetic field Andreev spectrum here is due to helical in-gap states, which in this situation are expected to be 2D surface Majorana fluids (distinct from the Majorana zero mode that has been detected in heterostructures).

\begin{figure}[]
\begin{center}
\includegraphics[width=16cm]{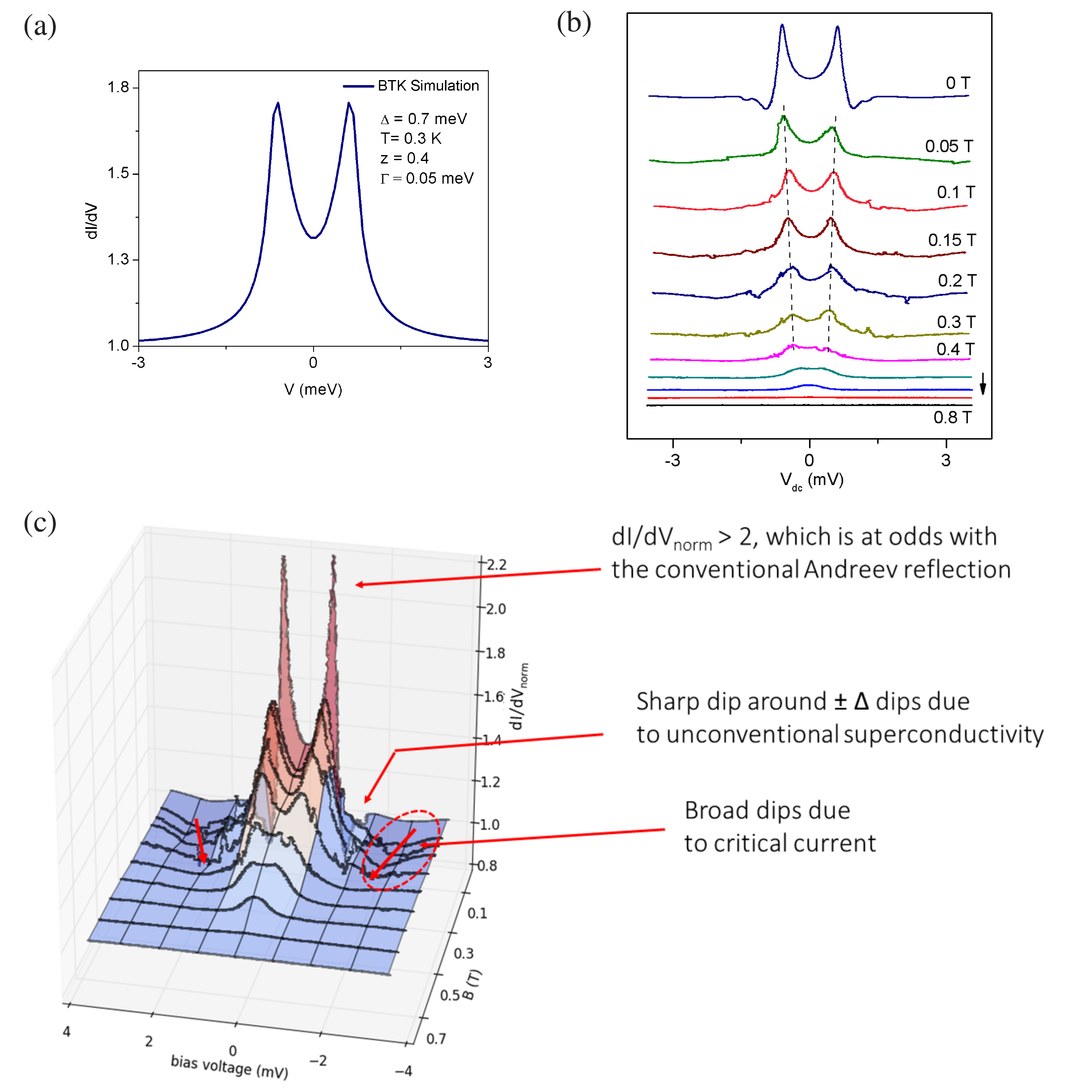}
\end{center}
\caption{(a) the simulation for the following parameter, similar to actual experimental values: superconducting gap, $\Delta$ = 0.7 meV, barrier strength, z = 0.4, temperature, T = 0.3 K, and quasiparticle interference parameter $\Gamma$ = 0.05 meV  (b) d$I$/d$V$ spectrum at 300 mK  (c) 3D plot of the d$I$/d$V$ spectrum depicting the dramatic increase of d$I$/d$V$ at zero magnetic field.}
\end{figure}

\clearpage

\section*{S5. Absence of WHH orbital and Pauli Limiting Effect}
Backscattering by impurities, even non-magnetic ones, suppresses odd-parity superconductivity as Anderson theorem does not hold \cite{fay1980coexistence, foulkes1977p}. Thus, we first check if the mean free path is greater than the coherent length. As shown in Fig. 4 in the main text, the upper critical field $B_{c2}$ values at different temperatures below $T_c$ is plotted and extrapolated to T = 0 by using the Ginzburg Landau (G-L) theory. $B_{c2}(0)$ is 0.69 T for $\beta$-PdBi$_2$ and a higher value of 0.89 T for K-doped crystal even with lower $T_c$. Using $B_{c2} = \Phi_0/2\pi\xi_{c2}^2$, where $\xi_{c2}$ is the Ginzburg-Landau coherence length and $\Phi_0$ is the flux quantum, we obtain $\xi$ = 19 nm for K-doped $\beta$-PdBi$_2$.  Assuming a spherical Fermi surface, we have wavenumber $k_F = (3\pi^2n)^{1/3}$ and with n = 4.81 x 10$^{27}$ m$^{-3}$ derived from the linear part of $\rho_{xy}$ for K doped $\beta$-PdBi$_2$, we find $l$ = 75 nm. This $l>>\xi$ combines contribution from both the surface and bulk state. If we use the bulk density, n = 3.4 x 10$^{28}$ m$^{-3}$, derived from S2, we have $l$ = 22 nm, which is still greater than $\xi$ = 19 nm. The doped crystal is sufficiently pure for odd-parity superconductivity. 

Under the BCS theory, superconductivity can be limited by orbital and spin effect of magnetic field. The orbital depairing effect is described by the WHH theory while the spin limiting effect is described by the Pauli paramagnetism formalism by equating the paramagnetic polarization energy to the SC condensation energy $\chi_n B_p^2 = N[0]\Delta^2$, where N[0] is the density of state, $\Delta$ is the SC gap, and from which the polarization field, $B_{c2}^{p}(0)$ = 1.86 T$_c$, is obtained. Under WHH theory in the clean limit, $B_{c2}^{orb}(0) = 0.72 T_c|dB_{c2}/dT|_{T_c}$ = 0.75 T for the doped sample. This is below the experimental $B_{c2}$ value, suggesting that superconductivity is not orbital-limited.  The spin limiting effect is described by the Pauli paramagnetism, $B_{c2}^{p}(0)$ = 1.86 T$_c$ = 8.184 T, which is way above the experimental B$_c$$_2$. So we have the relation: $B_{c2}^{orb}(0) < B_{c2}(0) << B_{c2}^{p}(0)$, a relation which is also observed in Cu$_x$Bi$_2$Se$_3$ \cite{bay2012superconductivity}. Now, when both the orbital and Pauli limiting effects are present, then $B_{c2} =B_{c2}^{orb}(0)/ \sqrt[]{1 + \alpha^2} $ = 0.74 T. $\alpha$ here is the Maki parameter \cite{maki1966effect};  $\alpha$= $\sqrt[]{2}B_{c2}^{orb}(0)/ B_{c2}^{p}(0)$ = 0.13. The expected theoretical $B_{c2}$ in the presence of both the orbital and spin limiting effects is lower than the experimental value 0.89 T. We can possibly conclude that the Pauli limiting effect is also absent.

\putbib

\end{bibunit}


\begin{thebibliography}{47}%
\makeatletter
\providecommand \@ifxundefined [1]{%
 \@ifx{#1\undefined}
}%
\providecommand \@ifnum [1]{%
 \ifnum #1\expandafter \@firstoftwo
 \else \expandafter \@secondoftwo
 \fi
}%
\providecommand \@ifx [1]{%
 \ifx #1\expandafter \@firstoftwo
 \else \expandafter \@secondoftwo
 \fi
}%
\providecommand \natexlab [1]{#1}%
\providecommand \enquote  [1]{``#1''}%
\providecommand \bibnamefont  [1]{#1}%
\providecommand \bibfnamefont [1]{#1}%
\providecommand \citenamefont [1]{#1}%
\providecommand \href@noop [0]{\@secondoftwo}%
\providecommand \href [0]{\begingroup \@sanitize@url \@href}%
\providecommand \@href[1]{\@@startlink{#1}\@@href}%
\providecommand \@@href[1]{\endgroup#1\@@endlink}%
\providecommand \@sanitize@url [0]{\catcode `\\12\catcode `\$12\catcode
  `\&12\catcode `\#12\catcode `\^12\catcode `\_12\catcode `\%12\relax}%
\providecommand \@@startlink[1]{}%
\providecommand \@@endlink[0]{}%
\providecommand \url  [0]{\begingroup\@sanitize@url \@url }%
\providecommand \@url [1]{\endgroup\@href {#1}{\urlprefix }}%
\providecommand \urlprefix  [0]{URL }%
\providecommand \Eprint [0]{\href }%
\providecommand \doibase [0]{http://dx.doi.org/}%
\providecommand \selectlanguage [0]{\@gobble}%
\providecommand \bibinfo  [0]{\@secondoftwo}%
\providecommand \bibfield  [0]{\@secondoftwo}%
\providecommand \translation [1]{[#1]}%
\providecommand \BibitemOpen [0]{}%
\providecommand \bibitemStop [0]{}%
\providecommand \bibitemNoStop [0]{.\EOS\space}%
\providecommand \EOS [0]{\spacefactor3000\relax}%
\providecommand \BibitemShut  [1]{\csname bibitem#1\endcsname}%
\let\auto@bib@innerbib\@empty
\bibitem [{\citenamefont {Moore}(2010)}]{moore2010birth}%
  \BibitemOpen
  \bibfield  {author} {\bibinfo {author} {\bibfnamefont {J.~E.}\ \bibnamefont
  {Moore}},\ }\href@noop {} {\bibfield  {journal} {\bibinfo  {journal}
  {Nature}\ }\textbf {\bibinfo {volume} {464}},\ \bibinfo {pages} {194}
  (\bibinfo {year} {2010})}\BibitemShut {NoStop}%
\bibitem [{\citenamefont {Schnyder}\ \emph {et~al.}(2008)\citenamefont
  {Schnyder}, \citenamefont {Ryu}, \citenamefont {Furusaki},\ and\
  \citenamefont {Ludwig}}]{schnyder2008classification}%
  \BibitemOpen
  \bibfield  {author} {\bibinfo {author} {\bibfnamefont {A.~P.}\ \bibnamefont
  {Schnyder}}, \bibinfo {author} {\bibfnamefont {S.}~\bibnamefont {Ryu}},
  \bibinfo {author} {\bibfnamefont {A.}~\bibnamefont {Furusaki}}, \ and\
  \bibinfo {author} {\bibfnamefont {A.~W.}\ \bibnamefont {Ludwig}},\
  }\href@noop {} {\bibfield  {journal} {\bibinfo  {journal} {Phys. Rev. B}\
  }\textbf {\bibinfo {volume} {78}},\ \bibinfo {pages} {195125} (\bibinfo
  {year} {2008})}\BibitemShut {NoStop}%
\bibitem [{\citenamefont {Hasan}\ and\ \citenamefont
  {Kane}(2010)}]{hasan2010colloquium}%
  \BibitemOpen
  \bibfield  {author} {\bibinfo {author} {\bibfnamefont {M.~Z.}\ \bibnamefont
  {Hasan}}\ and\ \bibinfo {author} {\bibfnamefont {C.~L.}\ \bibnamefont
  {Kane}},\ }\href@noop {} {\bibfield  {journal} {\bibinfo  {journal} {Rev.
  Mod. Phys.}\ }\textbf {\bibinfo {volume} {82}},\ \bibinfo {pages} {3045}
  (\bibinfo {year} {2010})}\BibitemShut {NoStop}%
\bibitem [{\citenamefont {Fu}\ and\ \citenamefont
  {Kane}(2007)}]{fu2007topological}%
  \BibitemOpen
  \bibfield  {author} {\bibinfo {author} {\bibfnamefont {L.}~\bibnamefont
  {Fu}}\ and\ \bibinfo {author} {\bibfnamefont {C.~L.}\ \bibnamefont {Kane}},\
  }\href@noop {} {\bibfield  {journal} {\bibinfo  {journal} {Phys. Rev. B}\
  }\textbf {\bibinfo {volume} {76}},\ \bibinfo {pages} {045302} (\bibinfo
  {year} {2007})}\BibitemShut {NoStop}%
\bibitem [{\citenamefont {Chen}\ \emph {et~al.}(2009)\citenamefont {Chen},
  \citenamefont {Analytis}, \citenamefont {Chu}, \citenamefont {Liu},
  \citenamefont {Mo}, \citenamefont {Qi}, \citenamefont {Zhang}, \citenamefont
  {Lu}, \citenamefont {Dai}, \citenamefont {Fang} \emph
  {et~al.}}]{chen2009experimental}%
  \BibitemOpen
  \bibfield  {author} {\bibinfo {author} {\bibfnamefont {Y.}~\bibnamefont
  {Chen}}, \bibinfo {author} {\bibfnamefont {J.~G.}\ \bibnamefont {Analytis}},
  \bibinfo {author} {\bibfnamefont {J.-H.}\ \bibnamefont {Chu}}, \bibinfo
  {author} {\bibfnamefont {Z.}~\bibnamefont {Liu}}, \bibinfo {author}
  {\bibfnamefont {S.-K.}\ \bibnamefont {Mo}}, \bibinfo {author} {\bibfnamefont
  {X.-L.}\ \bibnamefont {Qi}}, \bibinfo {author} {\bibfnamefont
  {H.}~\bibnamefont {Zhang}}, \bibinfo {author} {\bibfnamefont
  {D.}~\bibnamefont {Lu}}, \bibinfo {author} {\bibfnamefont {X.}~\bibnamefont
  {Dai}}, \bibinfo {author} {\bibfnamefont {Z.}~\bibnamefont {Fang}},  \emph
  {et~al.},\ }\href@noop {} {\bibfield  {journal} {\bibinfo  {journal}
  {Science}\ }\textbf {\bibinfo {volume} {325}},\ \bibinfo {pages} {178}
  (\bibinfo {year} {2009})}\BibitemShut {NoStop}%
\bibitem [{\citenamefont {Qi}\ \emph {et~al.}(2009)\citenamefont {Qi},
  \citenamefont {Hughes}, \citenamefont {Raghu},\ and\ \citenamefont
  {Zhang}}]{qi2009time}%
  \BibitemOpen
  \bibfield  {author} {\bibinfo {author} {\bibfnamefont {X.-L.}\ \bibnamefont
  {Qi}}, \bibinfo {author} {\bibfnamefont {T.~L.}\ \bibnamefont {Hughes}},
  \bibinfo {author} {\bibfnamefont {S.}~\bibnamefont {Raghu}}, \ and\ \bibinfo
  {author} {\bibfnamefont {S.-C.}\ \bibnamefont {Zhang}},\ }\href@noop {}
  {\bibfield  {journal} {\bibinfo  {journal} {Phys. Rev. Lett.}\ }\textbf
  {\bibinfo {volume} {102}},\ \bibinfo {pages} {187001} (\bibinfo {year}
  {2009})}\BibitemShut {NoStop}%
\bibitem [{\citenamefont {Grover}\ \emph {et~al.}(2014)\citenamefont {Grover},
  \citenamefont {Sheng},\ and\ \citenamefont
  {Vishwanath}}]{grover2014emergent}%
  \BibitemOpen
  \bibfield  {author} {\bibinfo {author} {\bibfnamefont {T.}~\bibnamefont
  {Grover}}, \bibinfo {author} {\bibfnamefont {D.}~\bibnamefont {Sheng}}, \
  and\ \bibinfo {author} {\bibfnamefont {A.}~\bibnamefont {Vishwanath}},\
  }\href@noop {} {\bibfield  {journal} {\bibinfo  {journal} {Science}\ }\textbf
  {\bibinfo {volume} {344}},\ \bibinfo {pages} {280} (\bibinfo {year}
  {2014})}\BibitemShut {NoStop}%
\bibitem [{\citenamefont {Mourik}\ \emph {et~al.}(2012)\citenamefont {Mourik},
  \citenamefont {Zuo}, \citenamefont {Frolov}, \citenamefont {Plissard},
  \citenamefont {Bakkers},\ and\ \citenamefont
  {Kouwenhoven}}]{mourik2012signatures}%
  \BibitemOpen
  \bibfield  {author} {\bibinfo {author} {\bibfnamefont {V.}~\bibnamefont
  {Mourik}}, \bibinfo {author} {\bibfnamefont {K.}~\bibnamefont {Zuo}},
  \bibinfo {author} {\bibfnamefont {S.~M.}\ \bibnamefont {Frolov}}, \bibinfo
  {author} {\bibfnamefont {S.}~\bibnamefont {Plissard}}, \bibinfo {author}
  {\bibfnamefont {E.~P.}\ \bibnamefont {Bakkers}}, \ and\ \bibinfo {author}
  {\bibfnamefont {L.~P.}\ \bibnamefont {Kouwenhoven}},\ }\href@noop {}
  {\bibfield  {journal} {\bibinfo  {journal} {Science}\ }\textbf {\bibinfo
  {volume} {336}},\ \bibinfo {pages} {1003} (\bibinfo {year}
  {2012})}\BibitemShut {NoStop}%
\bibitem [{\citenamefont {Fu}\ and\ \citenamefont
  {Kane}(2008)}]{fu2008superconducting}%
  \BibitemOpen
  \bibfield  {author} {\bibinfo {author} {\bibfnamefont {L.}~\bibnamefont
  {Fu}}\ and\ \bibinfo {author} {\bibfnamefont {C.~L.}\ \bibnamefont {Kane}},\
  }\href@noop {} {\bibfield  {journal} {\bibinfo  {journal} {Phys. Rev. Lett.}\
  }\textbf {\bibinfo {volume} {100}},\ \bibinfo {pages} {096407} (\bibinfo
  {year} {2008})}\BibitemShut {NoStop}%
\bibitem [{\citenamefont {Kitaev}(2003)}]{kitaev2003fault}%
  \BibitemOpen
  \bibfield  {author} {\bibinfo {author} {\bibfnamefont {A.~Y.}\ \bibnamefont
  {Kitaev}},\ }\href@noop {} {\bibfield  {journal} {\bibinfo  {journal} {Ann.
  Phys.}\ }\textbf {\bibinfo {volume} {303}},\ \bibinfo {pages} {2} (\bibinfo
  {year} {2003})}\BibitemShut {NoStop}%
\bibitem [{\citenamefont {Chadov}\ \emph {et~al.}(2010)\citenamefont {Chadov},
  \citenamefont {Qi}, \citenamefont {K{\"u}bler}, \citenamefont {Fecher},
  \citenamefont {Felser},\ and\ \citenamefont {Zhang}}]{chadov2010tunable}%
  \BibitemOpen
  \bibfield  {author} {\bibinfo {author} {\bibfnamefont {S.}~\bibnamefont
  {Chadov}}, \bibinfo {author} {\bibfnamefont {X.}~\bibnamefont {Qi}}, \bibinfo
  {author} {\bibfnamefont {J.}~\bibnamefont {K{\"u}bler}}, \bibinfo {author}
  {\bibfnamefont {G.~H.}\ \bibnamefont {Fecher}}, \bibinfo {author}
  {\bibfnamefont {C.}~\bibnamefont {Felser}}, \ and\ \bibinfo {author}
  {\bibfnamefont {S.~C.}\ \bibnamefont {Zhang}},\ }\href@noop {} {\bibfield
  {journal} {\bibinfo  {journal} {Nat. Mater.}\ }\textbf {\bibinfo {volume}
  {9}},\ \bibinfo {pages} {541} (\bibinfo {year} {2010})}\BibitemShut {NoStop}%
\bibitem [{\citenamefont {Fu}\ and\ \citenamefont {Berg}(2010)}]{fu2010odd}%
  \BibitemOpen
  \bibfield  {author} {\bibinfo {author} {\bibfnamefont {L.}~\bibnamefont
  {Fu}}\ and\ \bibinfo {author} {\bibfnamefont {E.}~\bibnamefont {Berg}},\
  }\href@noop {} {\bibfield  {journal} {\bibinfo  {journal} {Phys. Rev. Lett.}\
  }\textbf {\bibinfo {volume} {105}},\ \bibinfo {pages} {097001} (\bibinfo
  {year} {2010})}\BibitemShut {NoStop}%
\bibitem [{\citenamefont {Sakano}\ \emph {et~al.}(2015)\citenamefont {Sakano},
  \citenamefont {Okawa}, \citenamefont {Kanou}, \citenamefont {Sanjo},
  \citenamefont {Okuda}, \citenamefont {Sasagawa},\ and\ \citenamefont
  {Ishizaka}}]{sakano2015topologically}%
  \BibitemOpen
  \bibfield  {author} {\bibinfo {author} {\bibfnamefont {M.}~\bibnamefont
  {Sakano}}, \bibinfo {author} {\bibfnamefont {K.}~\bibnamefont {Okawa}},
  \bibinfo {author} {\bibfnamefont {M.}~\bibnamefont {Kanou}}, \bibinfo
  {author} {\bibfnamefont {H.}~\bibnamefont {Sanjo}}, \bibinfo {author}
  {\bibfnamefont {T.}~\bibnamefont {Okuda}}, \bibinfo {author} {\bibfnamefont
  {T.}~\bibnamefont {Sasagawa}}, \ and\ \bibinfo {author} {\bibfnamefont
  {K.}~\bibnamefont {Ishizaka}},\ }\href@noop {} {\bibfield  {journal}
  {\bibinfo  {journal} {Nat. Commun.}\ }\textbf {\bibinfo {volume} {6}}
  (\bibinfo {year} {2015})}\BibitemShut {NoStop}%
\bibitem [{\citenamefont {Shein}\ and\ \citenamefont
  {Ivanovskii}(2013)}]{shein2013electronic}%
  \BibitemOpen
  \bibfield  {author} {\bibinfo {author} {\bibfnamefont {I.~R.}\ \bibnamefont
  {Shein}}\ and\ \bibinfo {author} {\bibfnamefont {A.~L.}\ \bibnamefont
  {Ivanovskii}},\ }\href@noop {} {\bibfield  {journal} {\bibinfo  {journal} {J.
  Supercond. Nov. Magn.}\ }\textbf {\bibinfo {volume} {26}},\ \bibinfo {pages}
  {1} (\bibinfo {year} {2013})}\BibitemShut {NoStop}%
\bibitem [{\citenamefont {Iwaya}\ \emph {et~al.}(2017)\citenamefont {Iwaya},
  \citenamefont {Kohsaka}, \citenamefont {Okawa}, \citenamefont {Machida},
  \citenamefont {Bahramy}, \citenamefont {Hanaguri},\ and\ \citenamefont
  {Sasagawa}}]{iwaya2017full}%
  \BibitemOpen
  \bibfield  {author} {\bibinfo {author} {\bibfnamefont {K.}~\bibnamefont
  {Iwaya}}, \bibinfo {author} {\bibfnamefont {Y.}~\bibnamefont {Kohsaka}},
  \bibinfo {author} {\bibfnamefont {K.}~\bibnamefont {Okawa}}, \bibinfo
  {author} {\bibfnamefont {T.}~\bibnamefont {Machida}}, \bibinfo {author}
  {\bibfnamefont {M.}~\bibnamefont {Bahramy}}, \bibinfo {author} {\bibfnamefont
  {T.}~\bibnamefont {Hanaguri}}, \ and\ \bibinfo {author} {\bibfnamefont
  {T.}~\bibnamefont {Sasagawa}},\ }\href@noop {} {\bibfield  {journal}
  {\bibinfo  {journal} {Nat. commun.}\ }\textbf {\bibinfo {volume} {8}},\
  \bibinfo {pages} {976} (\bibinfo {year} {2017})}\BibitemShut {NoStop}%
\bibitem [{dop()}]{dopant}%
  \BibitemOpen
  \href@noop {} {\bibinfo  {journal} {In addition to K, we have also doped with
  Na, Pd, Te; all of which are superconducting}\ }\BibitemShut {NoStop}%
\bibitem [{\citenamefont {Zhao}\ \emph {et~al.}(2015)\citenamefont {Zhao},
  \citenamefont {Lv}, \citenamefont {Xue}, \citenamefont {Zhu}, \citenamefont
  {Deng}, \citenamefont {Wu},\ and\ \citenamefont {Chu}}]{zhao2015chemical}%
  \BibitemOpen
\bibfield  {journal} {  }\bibfield  {author} {\bibinfo {author} {\bibfnamefont
  {K.}~\bibnamefont {Zhao}}, \bibinfo {author} {\bibfnamefont {B.}~\bibnamefont
  {Lv}}, \bibinfo {author} {\bibfnamefont {Y.-Y.}\ \bibnamefont {Xue}},
  \bibinfo {author} {\bibfnamefont {X.-Y.}\ \bibnamefont {Zhu}}, \bibinfo
  {author} {\bibfnamefont {L.}~\bibnamefont {Deng}}, \bibinfo {author}
  {\bibfnamefont {Z.}~\bibnamefont {Wu}}, \ and\ \bibinfo {author}
  {\bibfnamefont {C.}~\bibnamefont {Chu}},\ }\href@noop {} {\bibfield
  {journal} {\bibinfo  {journal} {Phys. Rev. B}\ }\textbf {\bibinfo {volume}
  {92}},\ \bibinfo {pages} {174404} (\bibinfo {year} {2015})}\BibitemShut
  {NoStop}%
\bibitem [{Sup()}]{Supplemental}%
  \BibitemOpen
  \href@noop {} {\bibinfo  {journal} {See Supplemental Material at [URL will be
  inserted by publisher] for supporting figures and information.}\
  }\BibitemShut {NoStop}%
\bibitem [{\citenamefont {Ando}(2013)}]{ando2013topological}%
  \BibitemOpen
\bibfield  {journal} {  }\bibfield  {author} {\bibinfo {author} {\bibfnamefont
  {Y.}~\bibnamefont {Ando}},\ }\href@noop {} {\bibfield  {journal} {\bibinfo
  {journal} {J.Phys. Soc. Jpn}\ }\textbf {\bibinfo {volume} {82}},\ \bibinfo
  {pages} {102001} (\bibinfo {year} {2013})}\BibitemShut {NoStop}%
\bibitem [{\citenamefont {Hikami}\ \emph {et~al.}(1980)\citenamefont {Hikami},
  \citenamefont {Larkin},\ and\ \citenamefont {Nagaoka}}]{hikami1980spin}%
  \BibitemOpen
  \bibfield  {author} {\bibinfo {author} {\bibfnamefont {S.}~\bibnamefont
  {Hikami}}, \bibinfo {author} {\bibfnamefont {A.~I.}\ \bibnamefont {Larkin}},
  \ and\ \bibinfo {author} {\bibfnamefont {Y.}~\bibnamefont {Nagaoka}},\
  }\href@noop {} {\bibfield  {journal} {\bibinfo  {journal} {Prog. Theor.
  Phys.}\ }\textbf {\bibinfo {volume} {63}},\ \bibinfo {pages} {707} (\bibinfo
  {year} {1980})}\BibitemShut {NoStop}%
\bibitem [{\citenamefont {Altshuler}\ \emph {et~al.}(1982)\citenamefont
  {Altshuler}, \citenamefont {Aronov},\ and\ \citenamefont
  {Khmelnitsky}}]{altshuler1982effects}%
  \BibitemOpen
  \bibfield  {author} {\bibinfo {author} {\bibfnamefont {B.~L.}\ \bibnamefont
  {Altshuler}}, \bibinfo {author} {\bibfnamefont {A.~G.}\ \bibnamefont
  {Aronov}}, \ and\ \bibinfo {author} {\bibfnamefont {D.~E.}\ \bibnamefont
  {Khmelnitsky}},\ }\href@noop {} {\bibfield  {journal} {\bibinfo  {journal}
  {J. Phys. C}\ }\textbf {\bibinfo {volume} {15}},\ \bibinfo {pages} {7367}
  (\bibinfo {year} {1982})}\BibitemShut {NoStop}%
\bibitem [{\citenamefont {Daghero}\ and\ \citenamefont
  {Gonnelli}(2010)}]{daghero2010probing}%
  \BibitemOpen
  \bibfield  {author} {\bibinfo {author} {\bibfnamefont {D.}~\bibnamefont
  {Daghero}}\ and\ \bibinfo {author} {\bibfnamefont {R.}~\bibnamefont
  {Gonnelli}},\ }\href@noop {} {\bibfield  {journal} {\bibinfo  {journal}
  {Supercond. Sci. Technol.}\ }\textbf {\bibinfo {volume} {23}},\ \bibinfo
  {pages} {043001} (\bibinfo {year} {2010})}\BibitemShut {NoStop}%
\bibitem [{\citenamefont {Blonder}\ \emph {et~al.}(1982)\citenamefont
  {Blonder}, \citenamefont {Tinkham},\ and\ \citenamefont
  {Klapwijk}}]{blonder1982transition}%
  \BibitemOpen
  \bibfield  {author} {\bibinfo {author} {\bibfnamefont {G.~E.}\ \bibnamefont
  {Blonder}}, \bibinfo {author} {\bibfnamefont {M.}~\bibnamefont {Tinkham}}, \
  and\ \bibinfo {author} {\bibfnamefont {T.~M.}\ \bibnamefont {Klapwijk}},\
  }\href@noop {} {\bibfield  {journal} {\bibinfo  {journal} {Phys. Rev. B}\
  }\textbf {\bibinfo {volume} {25}},\ \bibinfo {pages} {4515} (\bibinfo {year}
  {1982})}\BibitemShut {NoStop}%
\bibitem [{\citenamefont {Sheet}\ \emph {et~al.}(2004)\citenamefont {Sheet},
  \citenamefont {Mukhopadhyay},\ and\ \citenamefont
  {Raychaudhuri}}]{sheet2004role}%
  \BibitemOpen
  \bibfield  {author} {\bibinfo {author} {\bibfnamefont {G.}~\bibnamefont
  {Sheet}}, \bibinfo {author} {\bibfnamefont {S.}~\bibnamefont {Mukhopadhyay}},
  \ and\ \bibinfo {author} {\bibfnamefont {P.}~\bibnamefont {Raychaudhuri}},\
  }\href@noop {} {\bibfield  {journal} {\bibinfo  {journal} {Phys. Rev. B}\
  }\textbf {\bibinfo {volume} {69}},\ \bibinfo {pages} {134507} (\bibinfo
  {year} {2004})}\BibitemShut {NoStop}%
\bibitem [{\citenamefont {Kashiwaya}\ \emph {et~al.}(2014)\citenamefont
  {Kashiwaya}, \citenamefont {Kashiwaya}, \citenamefont {Saitoh}, \citenamefont
  {Mawatari},\ and\ \citenamefont {Tanaka}}]{kashiwaya2014tunneling}%
  \BibitemOpen
  \bibfield  {author} {\bibinfo {author} {\bibfnamefont {S.}~\bibnamefont
  {Kashiwaya}}, \bibinfo {author} {\bibfnamefont {H.}~\bibnamefont
  {Kashiwaya}}, \bibinfo {author} {\bibfnamefont {K.}~\bibnamefont {Saitoh}},
  \bibinfo {author} {\bibfnamefont {Y.}~\bibnamefont {Mawatari}}, \ and\
  \bibinfo {author} {\bibfnamefont {Y.}~\bibnamefont {Tanaka}},\ }\href@noop {}
  {\bibfield  {journal} {\bibinfo  {journal} {Physica, E, Low-dimens. Syst.
  Nanostruct.}\ }\textbf {\bibinfo {volume} {55}},\ \bibinfo {pages} {25}
  (\bibinfo {year} {2014})}\BibitemShut {NoStop}%
\bibitem [{\citenamefont {Sasaki}\ \emph {et~al.}(2011)\citenamefont {Sasaki},
  \citenamefont {Kriener}, \citenamefont {Segawa}, \citenamefont {Yada},
  \citenamefont {Tanaka}, \citenamefont {Sato},\ and\ \citenamefont
  {Ando}}]{sasaki2011topological}%
  \BibitemOpen
  \bibfield  {author} {\bibinfo {author} {\bibfnamefont {S.}~\bibnamefont
  {Sasaki}}, \bibinfo {author} {\bibfnamefont {M.}~\bibnamefont {Kriener}},
  \bibinfo {author} {\bibfnamefont {K.}~\bibnamefont {Segawa}}, \bibinfo
  {author} {\bibfnamefont {K.}~\bibnamefont {Yada}}, \bibinfo {author}
  {\bibfnamefont {Y.}~\bibnamefont {Tanaka}}, \bibinfo {author} {\bibfnamefont
  {M.}~\bibnamefont {Sato}}, \ and\ \bibinfo {author} {\bibfnamefont
  {Y.}~\bibnamefont {Ando}},\ }\href@noop {} {\bibfield  {journal} {\bibinfo
  {journal} {Phys. Rev. Lett.}\ }\textbf {\bibinfo {volume} {107}},\ \bibinfo
  {pages} {217001} (\bibinfo {year} {2011})}\BibitemShut {NoStop}%
\bibitem [{\citenamefont {Yamakage}\ \emph {et~al.}(2012)\citenamefont
  {Yamakage}, \citenamefont {Yada}, \citenamefont {Sato},\ and\ \citenamefont
  {Tanaka}}]{yamakage2012theory}%
  \BibitemOpen
  \bibfield  {author} {\bibinfo {author} {\bibfnamefont {A.}~\bibnamefont
  {Yamakage}}, \bibinfo {author} {\bibfnamefont {K.}~\bibnamefont {Yada}},
  \bibinfo {author} {\bibfnamefont {M.}~\bibnamefont {Sato}}, \ and\ \bibinfo
  {author} {\bibfnamefont {Y.}~\bibnamefont {Tanaka}},\ }\href@noop {}
  {\bibfield  {journal} {\bibinfo  {journal} {Phys. Rev. B}\ }\textbf {\bibinfo
  {volume} {85}},\ \bibinfo {pages} {180509} (\bibinfo {year}
  {2012})}\BibitemShut {NoStop}%
\bibitem [{\citenamefont {Sato}\ and\ \citenamefont
  {Ando}(2017)}]{sato2017topological}%
  \BibitemOpen
  \bibfield  {author} {\bibinfo {author} {\bibfnamefont {M.}~\bibnamefont
  {Sato}}\ and\ \bibinfo {author} {\bibfnamefont {Y.}~\bibnamefont {Ando}},\
  }\href@noop {} {\bibfield  {journal} {\bibinfo  {journal} {Rep. Prog. Phys}\
  }\textbf {\bibinfo {volume} {80}},\ \bibinfo {pages} {076501} (\bibinfo
  {year} {2017})}\BibitemShut {NoStop}%
\bibitem [{\citenamefont {Mizushima}\ \emph {et~al.}(2014)\citenamefont
  {Mizushima}, \citenamefont {Yamakage}, \citenamefont {Sato},\ and\
  \citenamefont {Tanaka}}]{mizushima2014dirac}%
  \BibitemOpen
  \bibfield  {author} {\bibinfo {author} {\bibfnamefont {T.}~\bibnamefont
  {Mizushima}}, \bibinfo {author} {\bibfnamefont {A.}~\bibnamefont {Yamakage}},
  \bibinfo {author} {\bibfnamefont {M.}~\bibnamefont {Sato}}, \ and\ \bibinfo
  {author} {\bibfnamefont {Y.}~\bibnamefont {Tanaka}},\ }\href@noop {}
  {\bibfield  {journal} {\bibinfo  {journal} {Phys. Rev. B}\ }\textbf {\bibinfo
  {volume} {90}},\ \bibinfo {pages} {184516} (\bibinfo {year}
  {2014})}\BibitemShut {NoStop}%
\bibitem [{\citenamefont {Xu}\ \emph {et~al.}(2014)\citenamefont {Xu},
  \citenamefont {Alidoust}, \citenamefont {Belopolski}, \citenamefont
  {Richardella}, \citenamefont {Liu}, \citenamefont {Neupane}, \citenamefont
  {Bian}, \citenamefont {Huang}, \citenamefont {Sankar}, \citenamefont {Fang}
  \emph {et~al.}}]{xu2014momentum}%
  \BibitemOpen
  \bibfield  {author} {\bibinfo {author} {\bibfnamefont {S.-Y.}\ \bibnamefont
  {Xu}}, \bibinfo {author} {\bibfnamefont {N.}~\bibnamefont {Alidoust}},
  \bibinfo {author} {\bibfnamefont {I.}~\bibnamefont {Belopolski}}, \bibinfo
  {author} {\bibfnamefont {A.}~\bibnamefont {Richardella}}, \bibinfo {author}
  {\bibfnamefont {C.}~\bibnamefont {Liu}}, \bibinfo {author} {\bibfnamefont
  {M.}~\bibnamefont {Neupane}}, \bibinfo {author} {\bibfnamefont
  {G.}~\bibnamefont {Bian}}, \bibinfo {author} {\bibfnamefont {S.-H.}\
  \bibnamefont {Huang}}, \bibinfo {author} {\bibfnamefont {R.}~\bibnamefont
  {Sankar}}, \bibinfo {author} {\bibfnamefont {C.}~\bibnamefont {Fang}},  \emph
  {et~al.},\ }\href@noop {} {\bibfield  {journal} {\bibinfo  {journal} {Nat.
  Phys.}\ }\textbf {\bibinfo {volume} {10}},\ \bibinfo {pages} {943–}
  (\bibinfo {year} {2014})}\BibitemShut {NoStop}%
\bibitem [{\citenamefont {Dynes}\ \emph {et~al.}(1978)\citenamefont {Dynes},
  \citenamefont {Narayanamurti},\ and\ \citenamefont
  {Garno}}]{dynes1978direct}%
  \BibitemOpen
  \bibfield  {author} {\bibinfo {author} {\bibfnamefont {R.~C.}\ \bibnamefont
  {Dynes}}, \bibinfo {author} {\bibfnamefont {V.}~\bibnamefont
  {Narayanamurti}}, \ and\ \bibinfo {author} {\bibfnamefont {J.~P.}\
  \bibnamefont {Garno}},\ }\href@noop {} {\bibfield  {journal} {\bibinfo
  {journal} {Phys. Rev. Lett.}\ }\textbf {\bibinfo {volume} {41}},\ \bibinfo
  {pages} {1509} (\bibinfo {year} {1978})}\BibitemShut {NoStop}%
\bibitem [{\citenamefont {Werthamer}\ \emph {et~al.}(1966)\citenamefont
  {Werthamer}, \citenamefont {Helfand},\ and\ \citenamefont
  {Hohenberg}}]{werthamer1966temperature}%
  \BibitemOpen
  \bibfield  {author} {\bibinfo {author} {\bibfnamefont {N.}~\bibnamefont
  {Werthamer}}, \bibinfo {author} {\bibfnamefont {E.}~\bibnamefont {Helfand}},
  \ and\ \bibinfo {author} {\bibfnamefont {P.}~\bibnamefont {Hohenberg}},\
  }\href@noop {} {\bibfield  {journal} {\bibinfo  {journal} {Phys. Rev.}\
  }\textbf {\bibinfo {volume} {147}},\ \bibinfo {pages} {295} (\bibinfo {year}
  {1966})}\BibitemShut {NoStop}%
\bibitem [{\citenamefont {Scharnberg}\ and\ \citenamefont
  {Klemm}(1980)}]{scharnberg1980p}%
  \BibitemOpen
  \bibfield  {author} {\bibinfo {author} {\bibfnamefont {K.}~\bibnamefont
  {Scharnberg}}\ and\ \bibinfo {author} {\bibfnamefont {R.}~\bibnamefont
  {Klemm}},\ }\href@noop {} {\bibfield  {journal} {\bibinfo  {journal} {Phys.
  Rev. B}\ }\textbf {\bibinfo {volume} {22}},\ \bibinfo {pages} {5233}
  (\bibinfo {year} {1980})}\BibitemShut {NoStop}%
\bibitem [{\citenamefont {Maki}(1966)}]{maki1966effect}%
  \BibitemOpen
  \bibfield  {author} {\bibinfo {author} {\bibfnamefont {K.}~\bibnamefont
  {Maki}},\ }\href@noop {} {\bibfield  {journal} {\bibinfo  {journal} {Phys.
  Rev.}\ }\textbf {\bibinfo {volume} {148}},\ \bibinfo {pages} {362} (\bibinfo
  {year} {1966})}\BibitemShut {NoStop}%
\bibitem [{\citenamefont {Das}\ \emph {et~al.}(2011)\citenamefont {Das},
  \citenamefont {Suzuki}, \citenamefont {Tachiki},\ and\ \citenamefont
  {Kadowaki}}]{das2011spin}%
  \BibitemOpen
  \bibfield  {author} {\bibinfo {author} {\bibfnamefont {P.}~\bibnamefont
  {Das}}, \bibinfo {author} {\bibfnamefont {Y.}~\bibnamefont {Suzuki}},
  \bibinfo {author} {\bibfnamefont {M.}~\bibnamefont {Tachiki}}, \ and\
  \bibinfo {author} {\bibfnamefont {K.}~\bibnamefont {Kadowaki}},\ }\href@noop
  {} {\bibfield  {journal} {\bibinfo  {journal} {Phys. Rev. B}\ }\textbf
  {\bibinfo {volume} {83}},\ \bibinfo {pages} {220513} (\bibinfo {year}
  {2011})}\BibitemShut {NoStop}%
\bibitem [{\citenamefont {Kacmarcik}\ \emph {et~al.}(2016)\citenamefont
  {Kacmarcik}, \citenamefont {Pribulova}, \citenamefont {Samuely},
  \citenamefont {Szabo}, \citenamefont {Cambel}, \citenamefont {Soltys},
  \citenamefont {Herrera}, \citenamefont {Suderow}, \citenamefont
  {Correa-Orellana}, \citenamefont {Prabhakaran} \emph {et~al.}}]{2016single}%
  \BibitemOpen
  \bibfield  {author} {\bibinfo {author} {\bibfnamefont {J.}~\bibnamefont
  {Kacmarcik}}, \bibinfo {author} {\bibfnamefont {Z.}~\bibnamefont
  {Pribulova}}, \bibinfo {author} {\bibfnamefont {T.}~\bibnamefont {Samuely}},
  \bibinfo {author} {\bibfnamefont {P.}~\bibnamefont {Szabo}}, \bibinfo
  {author} {\bibfnamefont {V.}~\bibnamefont {Cambel}}, \bibinfo {author}
  {\bibfnamefont {J.}~\bibnamefont {Soltys}}, \bibinfo {author} {\bibfnamefont
  {E.}~\bibnamefont {Herrera}}, \bibinfo {author} {\bibfnamefont
  {H.}~\bibnamefont {Suderow}}, \bibinfo {author} {\bibfnamefont
  {A.}~\bibnamefont {Correa-Orellana}}, \bibinfo {author} {\bibfnamefont
  {D.}~\bibnamefont {Prabhakaran}},  \emph {et~al.},\ }\href@noop {} {\bibfield
   {journal} {\bibinfo  {journal} {Phys. Rev. B}\ }\textbf {\bibinfo {volume}
  {93}},\ \bibinfo {pages} {144502} (\bibinfo {year} {2016})}\BibitemShut
  {NoStop}%
\bibitem [{\citenamefont {Herrera}\ \emph {et~al.}(2015)\citenamefont
  {Herrera}, \citenamefont {Guillam{\'o}n}, \citenamefont {Galvis},
  \citenamefont {Correa}, \citenamefont {Fente}, \citenamefont {Luccas},
  \citenamefont {Mompean}, \citenamefont {Garc{\'\i}a-Hern{\'a}ndez},
  \citenamefont {Vieira}, \citenamefont {Brison} \emph
  {et~al.}}]{herrera2015magnetic}%
  \BibitemOpen
  \bibfield  {author} {\bibinfo {author} {\bibfnamefont {E.}~\bibnamefont
  {Herrera}}, \bibinfo {author} {\bibfnamefont {I.}~\bibnamefont
  {Guillam{\'o}n}}, \bibinfo {author} {\bibfnamefont {J.~A.}\ \bibnamefont
  {Galvis}}, \bibinfo {author} {\bibfnamefont {A.}~\bibnamefont {Correa}},
  \bibinfo {author} {\bibfnamefont {A.}~\bibnamefont {Fente}}, \bibinfo
  {author} {\bibfnamefont {R.}~\bibnamefont {Luccas}}, \bibinfo {author}
  {\bibfnamefont {F.}~\bibnamefont {Mompean}}, \bibinfo {author} {\bibfnamefont
  {M.}~\bibnamefont {Garc{\'\i}a-Hern{\'a}ndez}}, \bibinfo {author}
  {\bibfnamefont {S.}~\bibnamefont {Vieira}}, \bibinfo {author} {\bibfnamefont
  {J.-P.}\ \bibnamefont {Brison}},  \emph {et~al.},\ }\href@noop {} {\bibfield
  {journal} {\bibinfo  {journal} {Phys. Rev. B}\ }\textbf {\bibinfo {volume}
  {92}},\ \bibinfo {pages} {054507} (\bibinfo {year} {2015})}\BibitemShut
  {NoStop}%
\bibitem [{\citenamefont {Che}\ \emph {et~al.}(2016)\citenamefont {Che},
  \citenamefont {Le}, \citenamefont {Xu}, \citenamefont {Xing}, \citenamefont
  {Shi}, \citenamefont {Xu},\ and\ \citenamefont {Lu}}]{che2016absence}%
  \BibitemOpen
  \bibfield  {author} {\bibinfo {author} {\bibfnamefont {L.}~\bibnamefont
  {Che}}, \bibinfo {author} {\bibfnamefont {T.}~\bibnamefont {Le}}, \bibinfo
  {author} {\bibfnamefont {C.}~\bibnamefont {Xu}}, \bibinfo {author}
  {\bibfnamefont {X.}~\bibnamefont {Xing}}, \bibinfo {author} {\bibfnamefont
  {Z.}~\bibnamefont {Shi}}, \bibinfo {author} {\bibfnamefont {X.}~\bibnamefont
  {Xu}}, \ and\ \bibinfo {author} {\bibfnamefont {X.}~\bibnamefont {Lu}},\
  }\href@noop {} {\bibfield  {journal} {\bibinfo  {journal} {Phys. Rev. B}\
  }\textbf {\bibinfo {volume} {94}},\ \bibinfo {pages} {024519} (\bibinfo
  {year} {2016})}\BibitemShut {NoStop}%
\bibitem [{\citenamefont {Lv}\ \emph {et~al.}(2017)\citenamefont {Lv},
  \citenamefont {Wang}, \citenamefont {Zhang}, \citenamefont {Ding},
  \citenamefont {Li}, \citenamefont {Wang}, \citenamefont {He}, \citenamefont
  {Song}, \citenamefont {Ma},\ and\ \citenamefont {Xue}}]{lv2017experimental}%
  \BibitemOpen
  \bibfield  {author} {\bibinfo {author} {\bibfnamefont {Y.-F.}\ \bibnamefont
  {Lv}}, \bibinfo {author} {\bibfnamefont {W.-L.}\ \bibnamefont {Wang}},
  \bibinfo {author} {\bibfnamefont {Y.-M.}\ \bibnamefont {Zhang}}, \bibinfo
  {author} {\bibfnamefont {H.}~\bibnamefont {Ding}}, \bibinfo {author}
  {\bibfnamefont {W.}~\bibnamefont {Li}}, \bibinfo {author} {\bibfnamefont
  {L.}~\bibnamefont {Wang}}, \bibinfo {author} {\bibfnamefont {K.}~\bibnamefont
  {He}}, \bibinfo {author} {\bibfnamefont {C.-L.}\ \bibnamefont {Song}},
  \bibinfo {author} {\bibfnamefont {X.-C.}\ \bibnamefont {Ma}}, \ and\ \bibinfo
  {author} {\bibfnamefont {Q.-K.}\ \bibnamefont {Xue}},\ }\href@noop {}
  {\bibfield  {journal} {\bibinfo  {journal} {Sci. Bull.}\ } (\bibinfo {year}
  {2017})}\BibitemShut {NoStop}%
\bibitem [{\citenamefont {Zhang}\ \emph {et~al.}(2014)\citenamefont {Zhang},
  \citenamefont {Liu}, \citenamefont {Luo}, \citenamefont {Freeman},\ and\
  \citenamefont {Zunger}}]{zhang2014hidden}%
  \BibitemOpen
  \bibfield  {author} {\bibinfo {author} {\bibfnamefont {X.}~\bibnamefont
  {Zhang}}, \bibinfo {author} {\bibfnamefont {Q.}~\bibnamefont {Liu}}, \bibinfo
  {author} {\bibfnamefont {J.-W.}\ \bibnamefont {Luo}}, \bibinfo {author}
  {\bibfnamefont {A.~J.}\ \bibnamefont {Freeman}}, \ and\ \bibinfo {author}
  {\bibfnamefont {A.}~\bibnamefont {Zunger}},\ }\href@noop {} {\bibfield
  {journal} {\bibinfo  {journal} {Nat. Phys.}\ }\textbf {\bibinfo {volume}
  {10}},\ \bibinfo {pages} {387} (\bibinfo {year} {2014})}\BibitemShut
  {NoStop}%
\bibitem [{\citenamefont {Riley}\ \emph {et~al.}(2014)\citenamefont {Riley},
  \citenamefont {Mazzola}, \citenamefont {Dendzik}, \citenamefont {Michiardi},
  \citenamefont {Takayama}, \citenamefont {Bawden}, \citenamefont
  {Graner{\o}d}, \citenamefont {Leandersson}, \citenamefont {Balasubramanian},
  \citenamefont {Hoesch} \emph {et~al.}}]{riley2014direct}%
  \BibitemOpen
  \bibfield  {author} {\bibinfo {author} {\bibfnamefont {J.~M.}\ \bibnamefont
  {Riley}}, \bibinfo {author} {\bibfnamefont {F.}~\bibnamefont {Mazzola}},
  \bibinfo {author} {\bibfnamefont {M.}~\bibnamefont {Dendzik}}, \bibinfo
  {author} {\bibfnamefont {M.}~\bibnamefont {Michiardi}}, \bibinfo {author}
  {\bibfnamefont {T.}~\bibnamefont {Takayama}}, \bibinfo {author}
  {\bibfnamefont {L.}~\bibnamefont {Bawden}}, \bibinfo {author} {\bibfnamefont
  {C.}~\bibnamefont {Graner{\o}d}}, \bibinfo {author} {\bibfnamefont
  {M.}~\bibnamefont {Leandersson}}, \bibinfo {author} {\bibfnamefont
  {T.}~\bibnamefont {Balasubramanian}}, \bibinfo {author} {\bibfnamefont
  {M.}~\bibnamefont {Hoesch}},  \emph {et~al.},\ }\href@noop {} {\bibfield
  {journal} {\bibinfo  {journal} {Nat. Phys.}\ }\textbf {\bibinfo {volume}
  {10}},\ \bibinfo {pages} {835} (\bibinfo {year} {2014})}\BibitemShut
  {NoStop}%
\bibitem [{\citenamefont {Brydon}\ \emph {et~al.}(2014)\citenamefont {Brydon},
  \citenamefont {Sarma}, \citenamefont {Hui},\ and\ \citenamefont
  {Sau}}]{brydon2014odd}%
  \BibitemOpen
  \bibfield  {author} {\bibinfo {author} {\bibfnamefont {P.}~\bibnamefont
  {Brydon}}, \bibinfo {author} {\bibfnamefont {S.~D.}\ \bibnamefont {Sarma}},
  \bibinfo {author} {\bibfnamefont {H.-Y.}\ \bibnamefont {Hui}}, \ and\
  \bibinfo {author} {\bibfnamefont {J.~D.}\ \bibnamefont {Sau}},\ }\href@noop
  {} {\bibfield  {journal} {\bibinfo  {journal} {Phys. Rev. B}\ }\textbf
  {\bibinfo {volume} {90}},\ \bibinfo {pages} {184512} (\bibinfo {year}
  {2014})}\BibitemShut {NoStop}%
\bibitem [{\citenamefont {Kozii}\ and\ \citenamefont
  {Fu}(2015)}]{kozii2015odd}%
  \BibitemOpen
  \bibfield  {author} {\bibinfo {author} {\bibfnamefont {V.}~\bibnamefont
  {Kozii}}\ and\ \bibinfo {author} {\bibfnamefont {L.}~\bibnamefont {Fu}},\
  }\href@noop {} {\bibfield  {journal} {\bibinfo  {journal} {Phys. Rev. Lett.}\
  }\textbf {\bibinfo {volume} {115}},\ \bibinfo {pages} {207002} (\bibinfo
  {year} {2015})}\BibitemShut {NoStop}%
\bibitem [{\citenamefont {Wang}\ \emph {et~al.}(2016)\citenamefont {Wang},
  \citenamefont {Cho}, \citenamefont {Hughes},\ and\ \citenamefont
  {Fradkin}}]{wang2016topological}%
  \BibitemOpen
  \bibfield  {author} {\bibinfo {author} {\bibfnamefont {Y.}~\bibnamefont
  {Wang}}, \bibinfo {author} {\bibfnamefont {G.~Y.}\ \bibnamefont {Cho}},
  \bibinfo {author} {\bibfnamefont {T.~L.}\ \bibnamefont {Hughes}}, \ and\
  \bibinfo {author} {\bibfnamefont {E.}~\bibnamefont {Fradkin}},\ }\href@noop
  {} {\bibfield  {journal} {\bibinfo  {journal} {Phys. Rev. B}\ }\textbf
  {\bibinfo {volume} {93}},\ \bibinfo {pages} {134512} (\bibinfo {year}
  {2016})}\BibitemShut {NoStop}%
\bibitem [{\citenamefont {Yonezawa}\ \emph {et~al.}(2017)\citenamefont
  {Yonezawa}, \citenamefont {Tajiri}, \citenamefont {Nakata}, \citenamefont
  {Nagai}, \citenamefont {Wang}, \citenamefont {Segawa}, \citenamefont {Ando},\
  and\ \citenamefont {Maeno}}]{yonezawa2017thermodynamic}%
  \BibitemOpen
  \bibfield  {author} {\bibinfo {author} {\bibfnamefont {S.}~\bibnamefont
  {Yonezawa}}, \bibinfo {author} {\bibfnamefont {K.}~\bibnamefont {Tajiri}},
  \bibinfo {author} {\bibfnamefont {S.}~\bibnamefont {Nakata}}, \bibinfo
  {author} {\bibfnamefont {Y.}~\bibnamefont {Nagai}}, \bibinfo {author}
  {\bibfnamefont {Z.}~\bibnamefont {Wang}}, \bibinfo {author} {\bibfnamefont
  {K.}~\bibnamefont {Segawa}}, \bibinfo {author} {\bibfnamefont
  {Y.}~\bibnamefont {Ando}}, \ and\ \bibinfo {author} {\bibfnamefont
  {Y.}~\bibnamefont {Maeno}},\ }\href@noop {} {\bibfield  {journal} {\bibinfo
  {journal} {Nat. Phys.}\ }\textbf {\bibinfo {volume} {13}},\ \bibinfo {pages}
  {123} (\bibinfo {year} {2017})}\BibitemShut {NoStop}%
\bibitem [{\citenamefont {Pan}\ \emph {et~al.}(2016)\citenamefont {Pan},
  \citenamefont {Nikitin}, \citenamefont {Araizi}, \citenamefont {Huang},
  \citenamefont {Matsushita}, \citenamefont {Naka},\ and\ \citenamefont
  {De~Visser}}]{pan2016rotational}%
  \BibitemOpen
  \bibfield  {author} {\bibinfo {author} {\bibfnamefont {Y.}~\bibnamefont
  {Pan}}, \bibinfo {author} {\bibfnamefont {A.}~\bibnamefont {Nikitin}},
  \bibinfo {author} {\bibfnamefont {G.}~\bibnamefont {Araizi}}, \bibinfo
  {author} {\bibfnamefont {Y.}~\bibnamefont {Huang}}, \bibinfo {author}
  {\bibfnamefont {Y.}~\bibnamefont {Matsushita}}, \bibinfo {author}
  {\bibfnamefont {T.}~\bibnamefont {Naka}}, \ and\ \bibinfo {author}
  {\bibfnamefont {A.}~\bibnamefont {De~Visser}},\ }\href@noop {} {\bibfield
  {journal} {\bibinfo  {journal} {Sci. Rep.}\ }\textbf {\bibinfo {volume}
  {6}},\ \bibinfo {pages} {28632} (\bibinfo {year} {2016})}\BibitemShut
  {NoStop}%
\bibitem [{\citenamefont {Asaba}\ \emph {et~al.}(2017)\citenamefont {Asaba},
  \citenamefont {Lawson}, \citenamefont {Tinsman}, \citenamefont {Chen},
  \citenamefont {Corbae}, \citenamefont {Li}, \citenamefont {Qiu},
  \citenamefont {Hor}, \citenamefont {Fu},\ and\ \citenamefont
  {Li}}]{asaba2017rotational}%
  \BibitemOpen
  \bibfield  {author} {\bibinfo {author} {\bibfnamefont {T.}~\bibnamefont
  {Asaba}}, \bibinfo {author} {\bibfnamefont {B.}~\bibnamefont {Lawson}},
  \bibinfo {author} {\bibfnamefont {C.}~\bibnamefont {Tinsman}}, \bibinfo
  {author} {\bibfnamefont {L.}~\bibnamefont {Chen}}, \bibinfo {author}
  {\bibfnamefont {P.}~\bibnamefont {Corbae}}, \bibinfo {author} {\bibfnamefont
  {G.}~\bibnamefont {Li}}, \bibinfo {author} {\bibfnamefont {Y.}~\bibnamefont
  {Qiu}}, \bibinfo {author} {\bibfnamefont {Y.~S.}\ \bibnamefont {Hor}},
  \bibinfo {author} {\bibfnamefont {L.}~\bibnamefont {Fu}}, \ and\ \bibinfo
  {author} {\bibfnamefont {L.}~\bibnamefont {Li}},\ }\href@noop {} {\bibfield
  {journal} {\bibinfo  {journal} {Phys. Rev. X}\ }\textbf {\bibinfo {volume}
  {7}},\ \bibinfo {pages} {011009} (\bibinfo {year} {2017})}\BibitemShut
  {NoStop}%
\end{thebibliography}%


\begin{thebibliography}{9}%
\makeatletter
\providecommand \@ifxundefined [1]{%
 \@ifx{#1\undefined}
}%
\providecommand \@ifnum [1]{%
 \ifnum #1\expandafter \@firstoftwo
 \else \expandafter \@secondoftwo
 \fi
}%
\providecommand \@ifx [1]{%
 \ifx #1\expandafter \@firstoftwo
 \else \expandafter \@secondoftwo
 \fi
}%
\providecommand \natexlab [1]{#1}%
\providecommand \enquote  [1]{``#1''}%
\providecommand \bibnamefont  [1]{#1}%
\providecommand \bibfnamefont [1]{#1}%
\providecommand \citenamefont [1]{#1}%
\providecommand \href@noop [0]{\@secondoftwo}%
\providecommand \href [0]{\begingroup \@sanitize@url \@href}%
\providecommand \@href[1]{\@@startlink{#1}\@@href}%
\providecommand \@@href[1]{\endgroup#1\@@endlink}%
\providecommand \@sanitize@url [0]{\catcode `\\12\catcode `\$12\catcode
  `\&12\catcode `\#12\catcode `\^12\catcode `\_12\catcode `\%12\relax}%
\providecommand \@@startlink[1]{}%
\providecommand \@@endlink[0]{}%
\providecommand \url  [0]{\begingroup\@sanitize@url \@url }%
\providecommand \@url [1]{\endgroup\@href {#1}{\urlprefix }}%
\providecommand \urlprefix  [0]{URL }%
\providecommand \Eprint [0]{\href }%
\providecommand \doibase [0]{http://dx.doi.org/}%
\providecommand \selectlanguage [0]{\@gobble}%
\providecommand \bibinfo  [0]{\@secondoftwo}%
\providecommand \bibfield  [0]{\@secondoftwo}%
\providecommand \translation [1]{[#1]}%
\providecommand \BibitemOpen [0]{}%
\providecommand \bibitemStop [0]{}%
\providecommand \bibitemNoStop [0]{.\EOS\space}%
\providecommand \EOS [0]{\spacefactor3000\relax}%
\providecommand \BibitemShut  [1]{\csname bibitem#1\endcsname}%
\let\auto@bib@innerbib\@empty
\bibitem [{\citenamefont {Ando}(2013)}]{ando2013topological}%
  \BibitemOpen
  \bibfield  {author} {\bibinfo {author} {\bibfnamefont {Y.}~\bibnamefont
  {Ando}},\ }\href@noop {} {\bibfield  {journal} {\bibinfo  {journal} {J.Phys.
  Soc. Jpn}\ }\textbf {\bibinfo {volume} {82}},\ \bibinfo {pages} {102001}
  (\bibinfo {year} {2013})}\BibitemShut {NoStop}%
\bibitem [{\citenamefont {Qu}\ \emph {et~al.}(2010)\citenamefont {Qu},
  \citenamefont {Hor}, \citenamefont {Xiong}, \citenamefont {Cava},\ and\
  \citenamefont {Ong}}]{qu2010quantum}%
  \BibitemOpen
  \bibfield  {author} {\bibinfo {author} {\bibfnamefont {D.-X.}\ \bibnamefont
  {Qu}}, \bibinfo {author} {\bibfnamefont {Y.~S.}\ \bibnamefont {Hor}},
  \bibinfo {author} {\bibfnamefont {J.}~\bibnamefont {Xiong}}, \bibinfo
  {author} {\bibfnamefont {R.~J.}\ \bibnamefont {Cava}}, \ and\ \bibinfo
  {author} {\bibfnamefont {N.}~\bibnamefont {Ong}},\ }\href@noop {} {\bibfield
  {journal} {\bibinfo  {journal} {Science}\ }\textbf {\bibinfo {volume}
  {329}},\ \bibinfo {pages} {821} (\bibinfo {year} {2010})}\BibitemShut
  {NoStop}%
\bibitem [{\citenamefont {Ashcroft}\ and\ \citenamefont
  {Mermin}(1976)}]{ashcroft1976nd}%
  \BibitemOpen
  \bibfield  {author} {\bibinfo {author} {\bibfnamefont {N.~W.}\ \bibnamefont
  {Ashcroft}}\ and\ \bibinfo {author} {\bibfnamefont {D.~N.}\ \bibnamefont
  {Mermin}},\ }\href@noop {} {\enquote {\bibinfo {title} {Solid {State}
  {Physics}},}\ } (\bibinfo {year} {1976})\BibitemShut {NoStop}%
\bibitem [{\citenamefont {Daghero}\ and\ \citenamefont
  {Gonnelli}(2010)}]{daghero2010probing}%
  \BibitemOpen
  \bibfield  {author} {\bibinfo {author} {\bibfnamefont {D.}~\bibnamefont
  {Daghero}}\ and\ \bibinfo {author} {\bibfnamefont {R.}~\bibnamefont
  {Gonnelli}},\ }\href@noop {} {\bibfield  {journal} {\bibinfo  {journal}
  {Supercond. Sci. Technol.}\ }\textbf {\bibinfo {volume} {23}},\ \bibinfo
  {pages} {043001} (\bibinfo {year} {2010})}\BibitemShut {NoStop}%
\bibitem [{\citenamefont {Sasaki}\ \emph {et~al.}(2011)\citenamefont {Sasaki},
  \citenamefont {Kriener}, \citenamefont {Segawa}, \citenamefont {Yada},
  \citenamefont {Tanaka}, \citenamefont {Sato},\ and\ \citenamefont
  {Ando}}]{sasaki2011topological}%
  \BibitemOpen
  \bibfield  {author} {\bibinfo {author} {\bibfnamefont {S.}~\bibnamefont
  {Sasaki}}, \bibinfo {author} {\bibfnamefont {M.}~\bibnamefont {Kriener}},
  \bibinfo {author} {\bibfnamefont {K.}~\bibnamefont {Segawa}}, \bibinfo
  {author} {\bibfnamefont {K.}~\bibnamefont {Yada}}, \bibinfo {author}
  {\bibfnamefont {Y.}~\bibnamefont {Tanaka}}, \bibinfo {author} {\bibfnamefont
  {M.}~\bibnamefont {Sato}}, \ and\ \bibinfo {author} {\bibfnamefont
  {Y.}~\bibnamefont {Ando}},\ }\href@noop {} {\bibfield  {journal} {\bibinfo
  {journal} {Phys. Rev. Lett.}\ }\textbf {\bibinfo {volume} {107}},\ \bibinfo
  {pages} {217001} (\bibinfo {year} {2011})}\BibitemShut {NoStop}%
\bibitem [{\citenamefont {Fay}\ and\ \citenamefont
  {Appel}(1980)}]{fay1980coexistence}%
  \BibitemOpen
  \bibfield  {author} {\bibinfo {author} {\bibfnamefont {D.}~\bibnamefont
  {Fay}}\ and\ \bibinfo {author} {\bibfnamefont {J.}~\bibnamefont {Appel}},\
  }\href@noop {} {\bibfield  {journal} {\bibinfo  {journal} {Phys. Rev. B}\
  }\textbf {\bibinfo {volume} {22}},\ \bibinfo {pages} {3173} (\bibinfo {year}
  {1980})}\BibitemShut {NoStop}%
\bibitem [{\citenamefont {Foulkes}\ and\ \citenamefont
  {Gyorffy}(1977)}]{foulkes1977p}%
  \BibitemOpen
  \bibfield  {author} {\bibinfo {author} {\bibfnamefont {I.~F.}\ \bibnamefont
  {Foulkes}}\ and\ \bibinfo {author} {\bibfnamefont {B.}~\bibnamefont
  {Gyorffy}},\ }\href@noop {} {\bibfield  {journal} {\bibinfo  {journal} {Phys.
  Rev. B}\ }\textbf {\bibinfo {volume} {15}},\ \bibinfo {pages} {1395}
  (\bibinfo {year} {1977})}\BibitemShut {NoStop}%
\bibitem [{\citenamefont {Bay}\ \emph {et~al.}(2012)\citenamefont {Bay},
  \citenamefont {Naka}, \citenamefont {Huang}, \citenamefont {Luigjes},
  \citenamefont {Golden},\ and\ \citenamefont
  {De~Visser}}]{bay2012superconductivity}%
  \BibitemOpen
  \bibfield  {author} {\bibinfo {author} {\bibfnamefont {T.}~\bibnamefont
  {Bay}}, \bibinfo {author} {\bibfnamefont {T.}~\bibnamefont {Naka}}, \bibinfo
  {author} {\bibfnamefont {Y.}~\bibnamefont {Huang}}, \bibinfo {author}
  {\bibfnamefont {H.}~\bibnamefont {Luigjes}}, \bibinfo {author} {\bibfnamefont
  {M.}~\bibnamefont {Golden}}, \ and\ \bibinfo {author} {\bibfnamefont
  {A.}~\bibnamefont {De~Visser}},\ }\href@noop {} {\bibfield  {journal}
  {\bibinfo  {journal} {Phys. Rev. Lett.}\ }\textbf {\bibinfo {volume} {108}},\
  \bibinfo {pages} {057001} (\bibinfo {year} {2012})}\BibitemShut {NoStop}%
\bibitem [{\citenamefont {Maki}(1966)}]{maki1966effect}%
  \BibitemOpen
  \bibfield  {author} {\bibinfo {author} {\bibfnamefont {K.}~\bibnamefont
  {Maki}},\ }\href@noop {} {\bibfield  {journal} {\bibinfo  {journal} {Phys.
  Rev.}\ }\textbf {\bibinfo {volume} {148}},\ \bibinfo {pages} {362} (\bibinfo
  {year} {1966})}\BibitemShut {NoStop}%
\end{thebibliography}%


\begin{thebibliography}{0}%
\makeatletter
\providecommand \@ifxundefined [1]{%
 \@ifx{#1\undefined}
}%
\providecommand \@ifnum [1]{%
 \ifnum #1\expandafter \@firstoftwo
 \else \expandafter \@secondoftwo
 \fi
}%
\providecommand \@ifx [1]{%
 \ifx #1\expandafter \@firstoftwo
 \else \expandafter \@secondoftwo
 \fi
}%
\providecommand \natexlab [1]{#1}%
\providecommand \enquote  [1]{``#1''}%
\providecommand \bibnamefont  [1]{#1}%
\providecommand \bibfnamefont [1]{#1}%
\providecommand \citenamefont [1]{#1}%
\providecommand \href@noop [0]{\@secondoftwo}%
\providecommand \href [0]{\begingroup \@sanitize@url \@href}%
\providecommand \@href[1]{\@@startlink{#1}\@@href}%
\providecommand \@@href[1]{\endgroup#1\@@endlink}%
\providecommand \@sanitize@url [0]{\catcode `\\12\catcode `\$12\catcode
  `\&12\catcode `\#12\catcode `\^12\catcode `\_12\catcode `\%12\relax}%
\providecommand \@@startlink[1]{}%
\providecommand \@@endlink[0]{}%
\providecommand \url  [0]{\begingroup\@sanitize@url \@url }%
\providecommand \@url [1]{\endgroup\@href {#1}{\urlprefix }}%
\providecommand \urlprefix  [0]{URL }%
\providecommand \Eprint [0]{\href }%
\providecommand \doibase [0]{http://dx.doi.org/}%
\providecommand \selectlanguage [0]{\@gobble}%
\providecommand \bibinfo  [0]{\@secondoftwo}%
\providecommand \bibfield  [0]{\@secondoftwo}%
\providecommand \translation [1]{[#1]}%
\providecommand \BibitemOpen [0]{}%
\providecommand \bibitemStop [0]{}%
\providecommand \bibitemNoStop [0]{.\EOS\space}%
\providecommand \EOS [0]{\spacefactor3000\relax}%
\providecommand \BibitemShut  [1]{\csname bibitem#1\endcsname}%
\let\auto@bib@innerbib\@empty
\end{thebibliography}%
\end{document}